\documentclass[a4paper,12pt]{article}
\pdfoutput=1 

\usepackage{jheppub} 
\usepackage[T1]{fontenc} 

\usepackage[latin1]{inputenc}
\usepackage[english]{babel}

\usepackage{amssymb,amsthm,cancel,hyperref,graphicx,xcolor}
\usepackage{picinpar,graphicx,xypic}
\usepackage[subnum]{cases}
\usepackage{booktabs}
\usepackage{mathrsfs}
\usepackage{amsfonts}
\usepackage{latexsym}
\usepackage{booktabs,graphicx,hyperref,epsfig,multirow}
\usepackage{bm}
\allowdisplaybreaks
\def\bea{\begin{align}}
\def\eea{\end{align}}
\def\beq{\begin{equation}}
\def\eeq{\end{equation}}
\def\ba{\begin{eqnarray}}
\def\ea{\end{eqnarray}}
\def\be{\begin{equation}}
\def\ee{\end{equation}}
\definecolor{darkgreen}{HTML}{008000}
\newcommand{\sss}{\scriptscriptstyle\rm}
\newcommand{\muf}{\mu_{\rm\sss F}}
\newcommand{\mur}{\mu_{\rm\sss R}}
\newcommand{\mtop}{m_{\rm\sss top}}

\newcommand{\abs}[1]{\left|\,#1\,\right|}

\newcommand{\Ord}{\mathcal{O}}

\newcommand{\as}{\alpha_s}
\newcommand{\barkt}{\bar{k}_{\sss T}}

\def\({\left(}
\def\){\right)}
\def\[{\left[}
\def\]{\right]}
\def    \hepph  #1 {{\tt hep-ph/#1}}
\def    \hepex  #1 {{\tt hep-ex/#1}}
\long\def\symbolfootnote[#1]#2{\begingroup%
\def\thefootnote{\fnsymbol{footnote}}\footnote[#1]{#2}\endgroup}

\numberwithin{equation}{section}
\def\lapprox{\lower .7ex\hbox{$\;\stackrel{\textstyle <}{\sim}\;$}}
\def\gapprox{\lower .7ex\hbox{$\;\stackrel{\textstyle >}{\sim}\;$}}
\newcommand{\apprlim}[1]{\underset{#1}{\sim}}
\renewcommand{\(}{\left(}
\renewcommand{\)}{\right)}

\newcommand{\MSbar}{$\overline{\rm MS}$}

\newcommand{\Ca}{C_{\rm\sss A}}
\newcommand{\Cf}{C_{\rm\sss F}}

\newcommand{\Gf}{G_{\sss F}}
\newcommand{\mH}{m_{\sss H}}

\newcommand{\kt}{k_{\rm\sss T}}
\newcommand{\pt}{p_{\rm\sss T}}

\graphicspath{{ImagesHRTM/}}
\begin{document}
\begin{flushleft}
\begin{figure}[h]
\includegraphics[width=.2\textwidth]{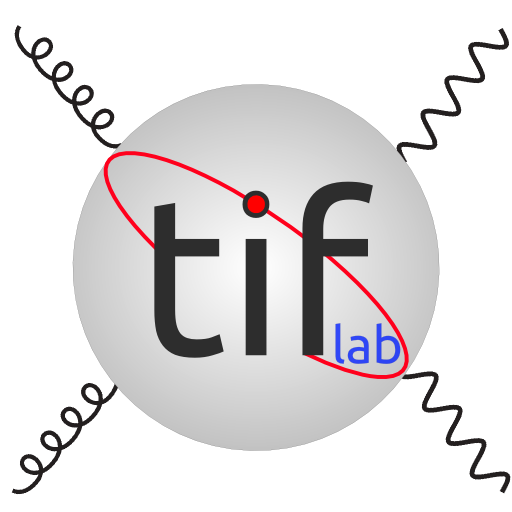}
\end{figure}
\end{flushleft}
\vspace{-5.0cm}
\begin{flushright}
TIF-UNIMI-2015-18
\end{flushright}

\vspace{2.0cm}

\begin{center}
{\Large \bf High energy resummation\\
of transverse momentum distributions:\\

\bigskip

Higgs in gluon fusion
}\\
\vspace{1.3cm}
Stefano Forte and Claudio Muselli\\
\vspace{.3cm}
{\it Tif Lab, Dipartimento di Fisica, Universit\`a di Milano and\\
INFN, Sezione di Milano,\\ Via Celoria 16, I-20133 Milano, Italy}\\
\vspace{2.0cm}
{\bf \large Abstract}
\end{center}
We derive a general resummation formula for transverse-momentum distributions
of hard processes at the leading logarithmic level in the high-energy
limit, to all orders in the strong coupling. Our result is based on a
suitable generalization of high-energy factorization theorems, whereby
all-order resummation is reduced to the determination of the
Born-level process but with incoming off-shell gluons. 
We validate our formula by applying it to  Higgs
production in gluon fusion in the infinite top mass limit. We 
check our result  up to next-to-leading order by comparison to
the high energy limit  of the exact expression and  to
next-to-next-to leading order by comparison to NNLL  transverse momentum
(Sudakov) resummation, and we predict  the high-energy 
behaviour  at next$^3$-to-leading order. We also show that the
structure of the result in the small transverse momentum limit agrees
to all orders
with general constraints from Sudakov resummation.

\clearpage
\tableofcontents
\clearpage

\section{High-energy  factorization}
\label{sec:intro}
High-energy resummation allows the computations of
contributions to hard  QCD processes, to all orders in the strong
coupling $\alpha_s$, which are enhanced by powers of logs of the
ratio $1/x$ of the center-of-mass energy $s$ to the scale of the hard
process $Q^2$: $x\equiv Q^2/s$. 
Like other resummation methods (such as Sudakov resummation) its value
is not only in enabling accurate phenomenology in kinematic regions in
which the resummed terms are large (i.e., in this case, when 
$\alpha_s \ln\frac{1}{x}\sim 1$), but also in providing information
on yet unknown higher order corrections. An interesting case in
point is the determination of the cross-section for Higgs in gluon
fusion, where high-energy resummation provided the first
information on the dependence of the cross-section on the top mass
beyond next-to-leading order, and the only available information at
N$^3$LO and beyond~\cite{Marzani:2008az,Harlander:2009my}.

High-energy resummation is based on factorization properties~\cite{Catani:1990xk,Catani:1990eg} which
have been known for a long time for total cross-sections, and,
originally applied to the photo- and
electro-production of heavy quarks, have been subsequently also derived for
deep-inelastic scattering~\cite{Catani:1994sq},
heavy quark hadro-production~\cite{Ball:2001pq}, 
 Higgs production, both without~\cite{Hautmann:2002tu} and with top mass
dependence~\cite{Marzani:2008az}, Drell-Yan
production~\cite{Marzani:2008uh}, 
and prompt-photon production~\cite{Diana:2009xv}. More recently, in Ref.~\cite{Caola:2010kv},
high-energy factorization was also derived for rapidity distributions,
and applied there to Higgs production in gluon fusion, both in the
infinite-top mass limit, and with full top mass dependence.

It is the purpose of this paper to extend these factorization results,
and the ensuing resummation methodology, to transverse momentum
distributions. This is an especially interesting generalization of the
high-energy resummation methodology both for reasons of principle,  and
in view of specific phenomenological applications.

Standard high-energy
factorization reduces the problem of computing the
cross-section to all orders in the high-energy limit to the
determination of a Born cross-section with incoming off-shell
gluons. Hence, for instance, Higgs production in gluon fusion is determined
to all-orders in the high-energy limit at the leading log level by the
knowledge of the cross-section for leading-order Higgs production in
gluon fusion through a quark loop, but with the two incoming gluons off-shell.
The all-order resummed result is obtainedby combining this off-shell
cross-section with the
information contained in the  anomalous dimension which resums to
leading  log accuracy 
the effect of radiation from incoming legs.   
The main insight on which our
results are based is   that putting
the incoming gluons off-shell is also
sufficient to determine the all-order transverse momentum dependence
in the high-energy limit, 
even when the leading-order process with
on-shell partons has trivial kinematics and no transverse momentum
dependence, such as in the case of Higgs in gluon fusion. 

A relevant phenomenological application of our result is the
determination of the transverse momentum distribution for Higgs
production in gluon fusion with full dependence on the top mass.
This is an important observable because the dependence of the Higgs
couplings on the top mass is a sensitive probe of the standard model,
and possible physics beyond it. However, this dependence is small for  the total
cross-section~\cite{Dittmaier:2011ti}, and only sizable for the transverse momentum
distribution~\cite{Baur:1989cm}. 
The latter, however, is only known at leading nontrivial
order (while it is known up to NNLO in the limit in which the top mass
goes to infinity~\cite{Boughezal:2015dra}). Use of our methods will allow for a simple
determination of the top mass
dependence of the Higgs momentum distribution to all orders, albeit in
the high-energy limit: this will be done in a companion paper.

The plan of this paper is the following:
after a brief summary of the standard  high energy resummation for
inclusive cross section
in Sect.~\ref{sec:HFgeneral}, we present in Sect.~\ref{sec:HFTM} 
the general resummed formula for transverse
momentum distributions, for hadro-, lepto-  and photo-production. In 
Sect.~\ref{sec:Higgs} we then apply our  formalism to Higgs
production: we determine the all-order resummed result for the
transverse momentum distribution in the infinite top mass limit, 
we expand it out up to N$^3$LO, and we check explicitly that up to NLO
it agrees with known results.  A check on our result at NNLO can be
obtained comparing to NNLL transverse momentum resummation, which
also contrains its general structure:
the relation between high-energy and transverse momentum resummation
is discussed in Sect.~\ref{sec:ptResum}, and conclusions are drawn in Sect.~\ref{sec:conclusion}

\section{The ladder expansion}
\label{sec:HFgeneral}
We briefly review the derivation of high-energy
factorization in the leading logarithmic approximation (LL$x$) for
inclusive cross section, following the approach of
Ref.~\cite{Caola:2010kv} (see also Ref.~\cite{caolathesis}), 
which facilitates its generalization to less
inclusive observables. In comparison to the derivation of
Ref.~\cite{Caola:2010kv}, which was built starting from the
electroproduction case, we deal directly with hadro-production, which
is the case we are mostly interested in.

\label{subsec:Finclusive}
We consider the production process of a state $ \mathcal{S}$  in a
hadronic collision characterized by a hard scale $Q$. Specifically,
(without loss of generality of the subsequent argument) we consider  a gluon
initiated process, like Higgs production 
\beq
\label{eq:process}
g(p) + g(n) \to \mathcal{S} + X,
\eeq
where $g(p)$ and $g(n)$ are initial-state gluons with momentum $p$ and
$n$ respectively.

\begin{figure}[htb]
\label{fig:hardladder2}
\centering
\includegraphics[width=0.5\textwidth]{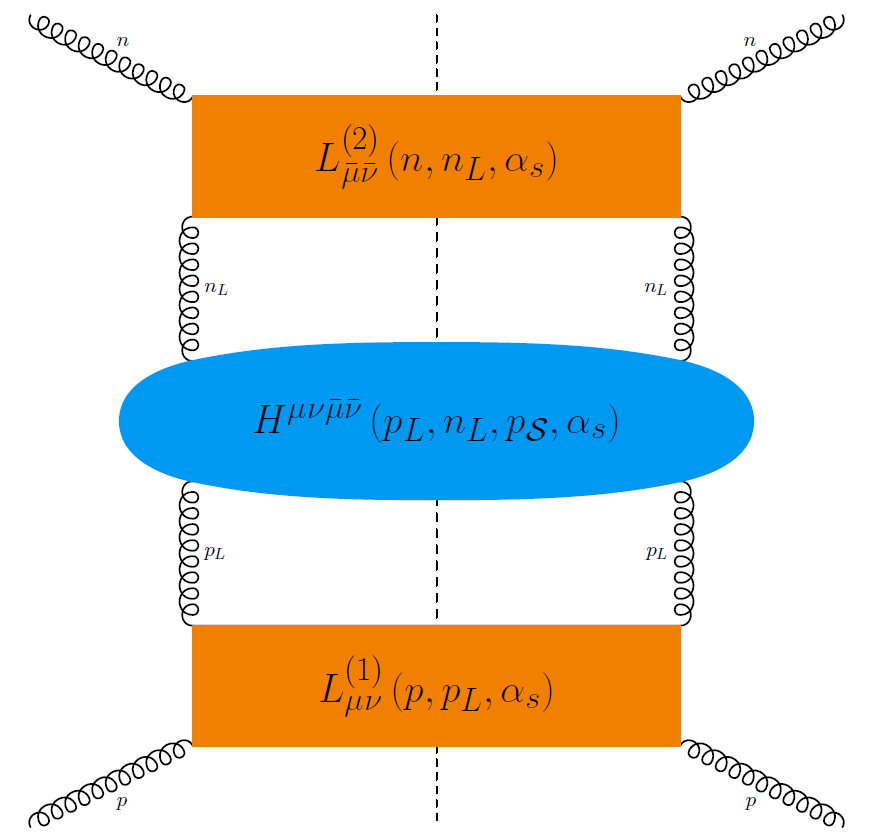}
\caption{Factorization of the partonic cross section in a hard part and two ladder parts}
\end{figure}

As in Refs.~\cite{Caola:2010kv,caolathesis}, we start from 
the observation~\cite{Ellis:1978ty,Catani:1990eg}
that in axial gauge the leading contribution in the 
high energy limit comes entirely  from cut diagrams which are at least
two-gluon-irreducible (2GI) in the $t$-channel, with radiation connecting the
two initial legs 
suppressed by powers of the center-of-mass energy $s$. It follows that
a (dimensionless) partonic cross section $\sigma$ can be written in
terms of  a process-dependent ``hard part'' $H^{\mu\nu\bar{\mu}\bar{\nu}}$, and  two
universal ``ladders'' $L_{\mu\nu}$:
\begin{align}
\label{eq:sigma}
\sigma\left(\frac{Q^2}{s},\frac{\muf^2}{Q^2},\frac{\mur^2}{Q^2}\right)=\int \frac{Q^2}{2s}\,&H^{\mu\nu\bar{\mu}\bar{\nu}}\(n_L,p_L,\Omega_\mathcal{S},\mur^2,\muf^2,\as\)\notag\\
&L_{\mu\nu}\(p_L,p,\mur^2,\muf^2,\as\)L_{\bar{\mu}\bar{\nu}}\(n_L,n,\mur^2,\muf^2,\as\) \left[d p_L\right]\left[d n_L\right],
\end{align}
where $Q^2$ is the hard scale of the process (typically the invariant
mass of $\mathcal{S}$),  $\Omega_\mathcal{S}$ denotes a set of variables which
characterize the kinematics of the final state $\mathcal{S}$, and $\left[d
  p_L\right]$ and $\left[d n_L\right]$ are  the integration measures over the
momenta connecting the hard part to the two ladders (see
Fig.~\ref{fig:hardladder2}).  In Eq.~\eqref{eq:sigma}
$\frac{1}{2s}$ is a flux factor, and the phase space is included in
the hard part, whence it can be removed if a differential
cross-section is sought.
The hard part and the ladders  are both separately symmetric under
exchange of the indices $\mu \leftrightarrow \nu $ and $\bar{\mu}
\leftrightarrow \bar{\nu}$. 

The hard part and the ladders  are both ultraviolet and collinear
divergent; renormalization and factorization then introduces a
dependence on the renormalization and factorization scales $\mur^2$ and
$\muf^2$. Because the running of the coupling is logarithmically
subleading (the coupling runs with the hard scale $Q^2$ and not with $s$), we can ignore
the $\mur^2$ dependence, which only goes through $\alpha_s(\mur^2)$ at
the  LL$x$ accuracy of our calculation. Furthermore,  in order to simplify
our derivation, we will assume that the hard part is two-particle irreducible, 
rather than just two-gluon
irreducible, in which case it is free of collinear
singularities~\cite{Ellis:1978ty,Curci:1980uw} and it is thus
independent of the factorization
scale. The extension to the case in which the hard part is two-particle
reducible and thus not collinear safe, such as deep-inelastic
scattering~\cite{Catani:1994sq} or Drell-Yan
production~\cite{Marzani:2008uh} is  nontrivial, but it does
not affect our argument, and it
will not be considered here.

The most general structure of the hard part and the ladders 
compatible with Lorentz invariance and the Ward identities is then
\begin{subequations}
\label{eq:HardLadder}
\begin{align}
H^{\mu\nu\bar{\mu}\bar{\nu}}\(n_L,p_L,\Omega_\mathcal{S},\as\)=&\(-g^{\mu\nu} + \frac{p_L^\mu p_L^\nu}{p_L^2}\)\(-g^{\bar{\mu}\bar{\nu}}+\frac{n_L^{\bar{\mu}} n_L^{\bar{\nu}}}{n_L^2}\)H_{\perp,\perp}\notag\\
+&\Bigg[n_L^2\(-g^{\mu \nu}+\frac{p_L^\mu p_L^\nu}{p_L^2}\)\(\frac{n_L^{\bar{\mu}}}{n_L^2}-\frac{p_L^{\bar{\mu}}}{\(n_L\cdot p_L\)}\)\(\frac{n_L^{\bar{\nu}}}{n_L^2}-\frac{p_L^{\bar{\nu}}}{\(n_L \cdot p_L\)}\)\notag\\
+&p_L^2\(\frac{p_L^\mu}{p_L^2}-\frac{n_L^\mu}{\(n_L\cdot p_L\)}\)\(\frac{p_L^\nu}{p_L^2}-\frac{n_L^\nu}{\(n_L \cdot p_L\)}\)\(-g^{\bar{\mu}\bar{\nu}}+\frac{n_L^{\bar{\mu}} n_L^{\bar{\nu}}}{n_L^2}\)\Bigg] H_{\perp,\parallel}\notag\\
+&p_L^2 n_L^2\(\frac{p_L^\mu}{p_L^2}-\frac{n_L^\mu}{\(n_L\cdot p_L\)}\)\(\frac{p_L^\nu}{p_L^2}-\frac{n_L^\nu}{\(n_L \cdot p_L\)}\) \notag\\
&\(\frac{n_L^{\bar{\mu}}}{n_L^2}-\frac{p_L^{\bar{\mu}}}{\(n_L\cdot p_L\)}\)\(\frac{n_L^{\bar{\nu}}}{n_L^2}-\frac{p_L^{\bar{\nu}}}{\(n_L \cdot p_L\)}\) H_{\parallel,\parallel}\notag\\
+&R^{\mu\nu\bar{\mu}\bar{\nu}} H_{\rm mixed}\label{eq:hmixed}\\
L^{\mu\nu}\(p_L,p,\muf^2,\as\)=&\frac{1}{p_L^2}\(-g^{\mu\nu} + \frac{p_L^\mu p_L^\nu}{p_L^2}\) L^{(1)}_{\perp} \notag\\
+&\(\frac{p_L^\mu}{p_L^2}-\frac{p^\mu}{\(p\cdot p_L\)}\)\(\frac{p_L^\nu}{p_L^2}-\frac{p^\nu}{\(p \cdot p_L\)}\)L^{(1)}_\parallel \\
L^{\bar{\mu}\bar{\nu}}\(n_L,n,\muf^2,\as\)=&\frac{1}{n_L^2}\(-g^{\bar{\mu}\bar{\nu}}+\frac{n_L^{\bar{\mu}} n_L^{\bar{\nu}}}{n_L^2}\) L^{(2)}_\perp \notag\\
+&\(\frac{n_L^{\bar{\mu}}}{n_L^2}-\frac{n^{\bar{\mu}}}{\(n \cdot n_L\)}\)\(\frac{n_L^{\bar{\nu}}}{n_L^2}-\frac{n^{\bar{\nu}}}{\(n\cdot n_L\)}\) L^{(2)}_\parallel,
\end{align}
\end{subequations}
in terms of  dimensionless scalar form factors 
\begin{subequations}
\label{eq:scalarfunction}
\begin{align}
H_{\rm mixed}=&H_{\rm mixed}\(\frac{Q^2}{\(n_L \cdot p_L\)},\frac{-p_L^2}{Q^2},\frac{-n_L^2}{Q^2},\Omega_\mathcal{S},\as\)\\
H_{\{ \perp , \parallel \},\{ \perp , \parallel \}}=&H_{\{ \perp , \parallel \},\{ \perp , \parallel \}}\(\frac{Q^2}{\(n_L \cdot p_L\)},\frac{-p_L^2}{Q^2},\frac{-n_L^2}{Q^2},\Omega_\mathcal{S},\as\)\\
L^{(1)}_{\{ \perp , \parallel\}}=&L^{(1)}_{\{\perp,\parallel\}}\(\frac{-p^2_L}{\(p\cdot p_L\)},\frac{\muf^2}{-p_L^2},\as\)\\
L^{(2)}_{\{\perp,\parallel\}}=&L^{(2)}_{\{\perp,\parallel\}}\(\frac{-n_L^2}{\(n\cdot n_L\)},\frac{\muf^2}{-n_L^2},\as\),
\end{align}
\end{subequations}
where with the notation $\{\perp,\parallel\}$ we mean that either of the
  two values can be chosen.
The tensor $R^{\mu\nu\bar{\mu}\bar{\nu}}$ contains all terms which mix
contribution coming from the two legs: it has a lengthy expression,
but it turns out to only require a single further  scalar form factor.

Equations~\eqref{eq:HardLadder} greatly simplify in the high energy
limit. 
In order to study it,  we define
\begin{equation}\label{eq:xdef}
x=\frac{Q^2}{s},
\end{equation}
and we introduce a Sudakov parametrization
for the two off-shell momenta $p_L$ and $n_L$: 
\begin{subequations}
\label{eq:sudakovparametrization}
\begin{align}
p_L=&z\,p - \mathbf{k}-\frac{\kt^2}{s\(1-z\)}\,n=\(\sqrt{\frac{s}{2}}\,z,-\frac{\kt^2}{\sqrt{2s}\(1-z\)},-\mathbf{k}_{\sss T}\)\\
n_L=&\bar{z}\,n - \mathbf{\bar{k}}-\frac{\barkt^2}{s\(1-\bar{z}\)}\,p=\(-\frac{\barkt^2}{\sqrt{2s}\(1-\bar{z}\)},\sqrt{\frac{s}{2}}\,\bar{z},-\mathbf{\bar{k}}_{\sss T}\),
\end{align}
\end{subequations}
where $\mathbf{k}$ and
$\mathbf{\bar{k}}$ are purely transverse spacelike  four-vectors
with $\mathbf{k}^2=-\kt^2 <0$ and $\mathbf{\bar{k}}^2=-\barkt^2<0$,
and $s=2\(p \cdot n\)$.  
With this parametrization, the integration measures $\left[dp_L\right]$ and $\left[dn_L\right]$ are
\begin{equation}
\left[dp_L\right]=\frac{dz}{2\(1-z\)}d^2\mathbf{k};\qquad \left[dn_L\right]=\frac{d\bar{z}}{2\(1-\bar{z}\)}d^2\mathbf{\bar{k}}.
\end{equation}

The high-energy limit is the limit in which $x\to0$: we wish to
determine the dominant power of $x$ contributing to $\sigma$,
Eq.~\eqref{eq:sigma}, with terms proportional to $\ln x$ 
included to all orders in $\alpha_s$ at the leading logarithmic
(LL$x$) level. We then
observe that, because  the integration over  $z$ and
$\bar{z}$ ranges from   $x$ to $1$, terms which are enhanced at small
$x$  come from the small $z$ and $\bar{z}$ region. The moduli of the transverse momenta  
$\kt^2$ and $\bar{\kt^2}$ are of order of the hard scale $Q^2$ which
bounds them from above, and thus in
the high energy regime $Q^2 \ll s$, they satisfy $\frac{\kt^2}{s} \ll
1$ and $\frac{\barkt^2}{s} \ll 1$. Therefore, the  high energy regime is
\begin{equation}\label{eq:helim}
z \ll 1,\qquad \frac{\kt^2}{s} \ll 1;\qquad\qquad 
\bar{z} \ll 1,\qquad \frac{\barkt^2}{s} \ll 1,
\end{equation} 
and subleading terms in $z$, $\bar{z}$, $\frac{\kt^2}{s}$ or
$\frac{\barkt^2}{s}$ upon integration lead to  power-suppressed
$\Ord\(x\)$ terms.

We can now simplify  Eq.~\eqref{eq:HardLadder}. First, we recall~\cite{Ellis:1978ty} that interference between
emissions from different legs is 
power-suppressed in $s$. It follows that $H_{\rm mixed}$
Eq.~\eqref{eq:hmixed} is subleading.
Furthermore,
we note~\cite{Caola:2010kv} that in the limit Eq.~\eqref{eq:helim} the
dependence of the remaining scalar functions simplifies:
\begin{align}\label{eq:scalhe}
H_{\{ \perp , \parallel \},\{ \perp , \parallel \}}\(\frac{Q^2}{\(n_L \cdot p_L\)},\frac{-p_L^2}{Q^2},\frac{-n_L^2}{Q^2},\Omega_\mathcal{S},\as\)=&H_{\{ \perp , \parallel \},\{ \perp , \parallel \}}\(\frac{x}{z\bar{z}},\frac{\kt^2}{Q^2},\frac{\barkt^2}{Q^2},\Omega_\mathcal{S},\as\)\(1+\Ord\(z,\bar{z}\)\)\\
L^{(1)}_{\{\perp,\parallel\}}\(\frac{-p^2_L}{\(p\cdot p_L\)},\frac{\muf^2}{-p_L^2},\as\)=&L^{(1)}_{\{\perp,\parallel\}}\(\frac{\muf^2}{\kt^2},\as\)\(1+\Ord\(z\)\)\\
L^{(2)}_{\{\perp,\parallel\}}\(\frac{-n_L^2}{\(n\cdot n_L\)},\frac{\muf^2}{-n_L^2},\as\)=&L^{(2)}_{\{\perp,\parallel\}}\(\frac{\muf^2}{\barkt^2},\as\)\(1+\Ord\(\bar{z}\)\)
\end{align}
up to terms that are suppressed by power of $z$ or $\bar{z}$. 
Finally, 
power counting arguments~\cite{Ellis:1978ty, Catani:1990eg} lead
to the conclusion that the transverse scalar functions
Eq.~\eqref{eq:scalhe} are no more
singular that the longitudinal ones: it follows
that the partonic cross
section, Eq.~\eqref{eq:sigma} in the small $x$ limit
has the form
\begin{align}
\label{eq:sigmahighenergy}
\sigma\left(x,\frac{\mu^2_F}{Q^2}\right)&=\int \left[\frac{x}{2 z \bar{z}} H_{\parallel,\parallel}\(\frac{x}{z\bar{z}},\frac{\kt^2}{Q^2},\frac{\barkt^2}{Q^2},\Omega_\mathcal{S},\as\)\right]\notag\\
&\;\left[2\pi L^{(1)}_{\parallel}\(\frac{\muf^2}{\kt^2},\as\)\right]\left[2\pi L^{(2)}_{\parallel}\(\frac{\muf^2}{\barkt^2},\as\)\right]\frac{dz}{z}\frac{d\bar{z}}{\bar{z}}\frac{d\kt^2}{\kt^2}\frac{d\barkt^2}{\barkt^2}\frac{d\theta}{2\pi}\frac{d\bar{\theta}}{2\pi}+\Ord\(z,\bar{z}\),
\end{align}
where $\theta$ and $\bar \theta$ are the azimuthal angles of the
transverse momenta $\mathbf{k}$ and  $\mathbf{\bar{k}}$, and at LL$x$
$\alpha_s$ is
fixed, and thus $\sigma$ is $\mu_R$-independent.

We note that the dependence on $\theta$ and $\bar \theta$ is entirely
contained in the hard part. Also,  in the  high-energy limit the
longitudinal projectors which carry the tensor structure of the term
proportional to $H_{\parallel,\parallel}$  Eq.~\eqref{eq:HardLadder}
reduce to~\cite{Catani:1990eg}
\begin{equation}
\label{eq:projd}
\mathcal{P}^{\mu\nu}=\frac{k^\mu k^\nu}{\kt^2};\qquad 
\mathcal{P}^{\bar{\mu}\bar{\nu}}=\frac{\bar{k}^{\bar{\mu}}
  \bar{k}^{\bar{\nu}}}{\barkt^2}.
\end{equation}
We can thus rewrite the cross-section Eq.~\eqref{eq:sigmahighenergy} in
terms of a generalized coefficient function
\begin{align}
C\(\frac{x}{z \bar{z}},\frac{\kt^2}{Q^2},\frac{\barkt^2}{Q^2},\as\)&\equiv
\int \frac{d\theta}{2\pi}\frac{d\bar{\theta}}{2\pi}\,\frac{x}{2 z
  \bar{z}}H_{\parallel,\parallel}\(\frac{x}{z
  \bar{z}},\frac{\kt^2}{Q^2},\frac{\barkt^2}{Q^2},\Omega_\mathcal{S},\as\)\nonumber\\
&=
\int \frac{d\theta}{2\pi}\frac{d\bar{\theta}}{2\pi}\,\frac{x}{2 z \bar{z}}\left[\mathcal{P}^{\mu\nu}\mathcal{P}^{\bar{\mu}\bar{\nu}}H_{\mu\nu\bar{\mu}\bar{\nu}}\right].
\label{eq:C}
\end{align}
The coefficient function Eq.~\eqref{eq:C}
is recognized as 
the cross section for the partonic process
\beq
g^*\(q\) + g^*\(r\) \to \mathcal{S}
\eeq
with two incoming off-shell gluon with momenta
\begin{align}
&q= z p + k  \qquad q^2=-\kt^2\\
&r=\bar{z} n + \bar{k} \qquad r^2=-\barkt^2,
\end{align} 
and the projectors Eq.~\eqref{eq:projd} viewed as polarization sums. 

Because the hard part is 2GI, the coefficient function is regular in
the $x\to 0$ limit, and  small $x$ singularities are only contained 
in the ladders. 
In Ref.~\cite{Catani:1990xk,Catani:1990eg} they are computed at LL$x$ level in terms
of a gluon Green function, which in turns sum leading logs of $x$ by
iterating a BFKL~\cite{Lipatov:polo} kernel. In
Ref.~\cite{Caola:2010kv}
 they were instead determined  using  the generalized ladder expansion 
of Ref.~\cite{Curci:1980uw}. This derivation is closer to that of
standard collinear factorization, and thus more suitable to applications
of high-energy resummation to standard, collinear-factorized hard
partonic cross-section, and specifically to its extension to less
inclusive quantities.

\begin{figure}[htb]
\label{fig:kernels}
\centering
\includegraphics[width=0.5\textwidth]{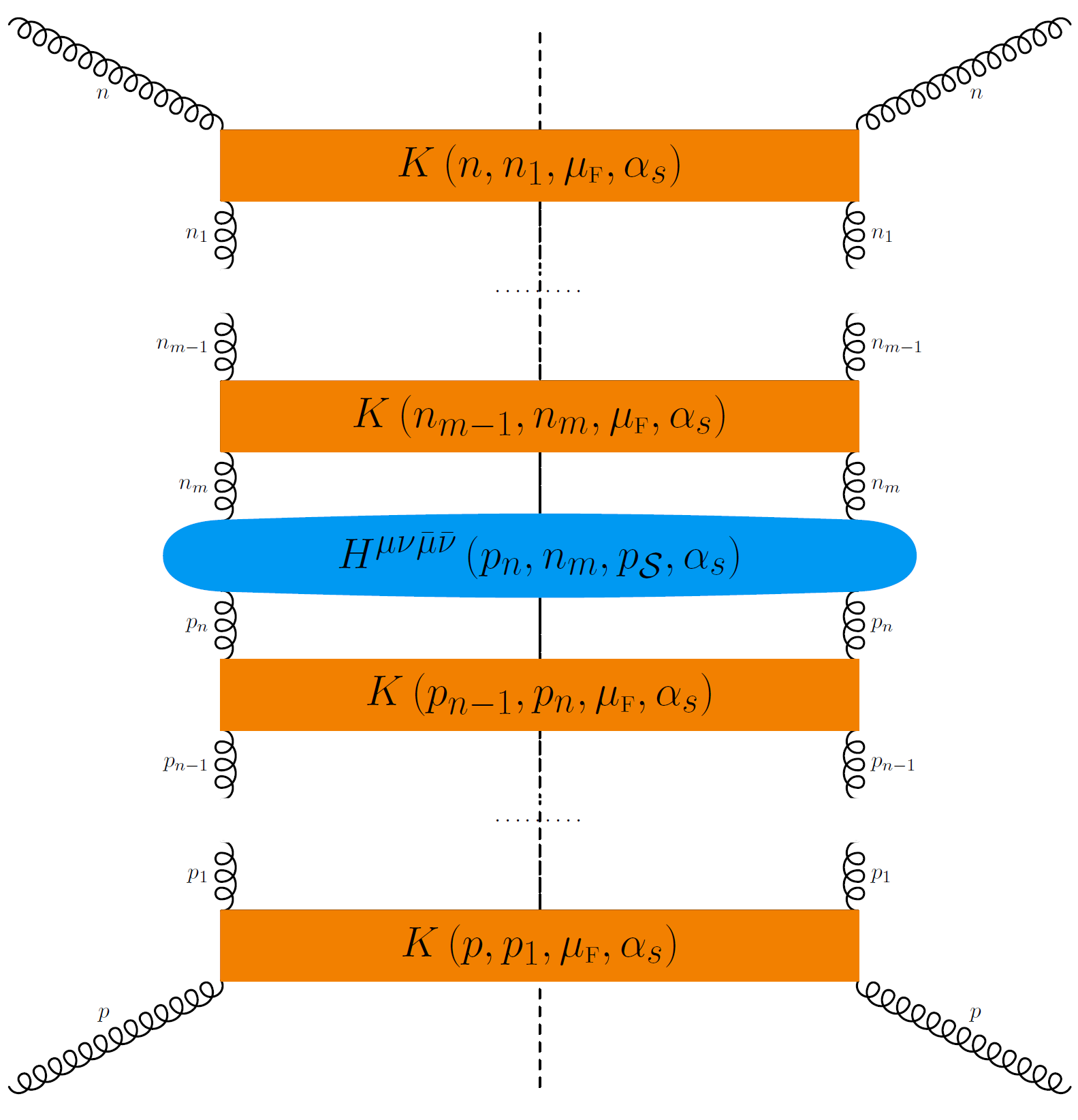}
\caption{Computation of the ladder parts by iterative insertion of the Kernel $K$}
\end{figure}

\label{subsec:GLE}
The ladders  contain collinear singularities that must be factorized
in
the parton distributions after regularization; this can be done in an
iterative way~\cite{Curci:1980uw} which also leads to small $x$
resummation, as explained 
in Ref.~\cite{Caola:2010kv}, which we follow in view of our desired
generalization. In this approach, the ladders $L^{(1)}_{\parallel}$
and $L^{(2)}_{\parallel}$ are obtained 
by iteration of a 2GI kernel $K\(p_i,p_{i-1},\mu,\as\)$ or
$K\(n_i,n_{i-1},\mu,\as\)$ with $i=1,2,\dots,n$, connected
 by a pair of $t$-channel gluons (see Fig.~\ref{fig:kernels}). The
 transverse  momenta of the gluons are ordered, ${\kt^2}\,_1 \ll
 {\kt^2}\,_2 \ll \dots \ll {\kt^2}\,_n = \kt^2$ (and ${\barkt^2}\,_1
 \ll {\barkt^2}\,_2 \ll \dots \ll {\barkt^2}\,_n =\barkt^2$), with
 small $x$ resummation performed by computing the kernels at LL$x$ 
to all orders
 in $\alpha_s$.

We start from a regularized version of the
expression  Eq.~\eqref{eq:sigmahighenergy} for the cross-section,
written in terms of the coefficient function $C$,
Eq.~\eqref{eq:C}:
\begin{align}
\label{eq:sigma1}
\sigma\(x,\frac{\muf^2}{Q^2},\as;\epsilon\)=\(Q^2\)^{2\epsilon}\int
&C\(\frac{x}{z
  \bar{z}},\frac{\kt^2}{Q^2},\frac{\barkt^2}{Q^2},\as;
\epsilon\)\left[2\pi L^{(1)}_{\parallel}\(z,
  \(\frac{\muf^2}{\kt^2}\)^\epsilon,\as; \epsilon\)\right]\notag\\ 
&\left[2\pi
  L^{(2)}_{\parallel}\(\bar{z},\(\frac{\muf^2}{\barkt^2}\)^\epsilon,\as;
  \epsilon\)\right]\frac{dz}{z}\frac{d\bar{z}}{\bar{z}}\frac{d\kt^2}{\(\kt^2\)^{1+\epsilon}}\frac{d\barkt^2}{\(\barkt^2\)^{1+\epsilon}},
\end{align}
where the dependence on $z$ and $\bar z$ in the ladders is $\Ord(\epsilon)$~\cite{Caola:2010kv}.
We factorize, as usual, the convolutions  by Mellin transformation
\begin{align}
\label{eq:sigma2}
\sigma\(N,\frac{\muf^2}{Q^2},\as;\epsilon\)=&\int_0^\infty \frac{d\xi}{\xi^{1+\epsilon}}\int_0^\infty \frac{d\bar{\xi}}{\bar{\xi}^{1+\epsilon}} C\(N,\xi,\bar{\xi},\as; \epsilon\)\notag\\
&\left[2\pi L^{(1)}_{\parallel}\(N, \(\frac{\muf^2}{Q^2\xi}\)^\epsilon,\as; \epsilon\)\right]\left[2\pi L^{(2)}_{\parallel}\(N,\(\frac{\muf^2}{Q^2 \bar{\xi}}\)^\epsilon,\as; \epsilon\)\right],
\end{align}
with
\begin{align}\label{eq:mellin}
f(N)=\int_0^1 dx\, x^{N-1} f(x);\qquad  f(x)=&\frac{1}{2\pi i}\int_{N_0-i\infty}^{N_0+i\infty}dN\,x^{-N} f(N),
\end{align}
where we have introduced dimensionless variables
\begin{equation}\label{eq:xidef}
\xi=\frac{\kt^2}{Q^2},\qquad \bar{\xi}=\frac{\barkt^2}{Q^2}.
\end{equation}
Note that the $Q^2$-dependence of the ladders is fictitious, as
$\frac{\muf^2}{Q^2\xi}=\frac{\muf^2}{\kt^2}$. Upon Mellin
transformation, powers of $\ln\frac{1}{x}$ are mapped onto poles at
$N=0$: note that the Mellin variable in Eq.~\eqref{eq:mellin}, as
usual in the context of high-energy resummation, is
shifted by one unit in comparison to the more customary definition.

The observation~\cite{Curci:1980uw} that collinear poles
 in $\epsilon$ are all produced by the integrations over the
 transverse momenta  ${\kt^2}$, $\barkt^2$  connecting the
 kernels leads to the identification of the kernel itself with  the 
anomalous dimension $\gamma$ in $d=4-2\epsilon$ dimensions, which in
our case must be computed to all orders in $\as$ to  LL$x$
accuracy~\cite{Caola:2010kv}:
\beq\label{eq:llxk}
K\(N,\(\frac{\muf^2}{Q^2{\xi}}\)^\epsilon,\as; \epsilon\)=\gamma\(N,\(\frac{\muf^2}{Q^2 \xi}\)^\epsilon,\as; \epsilon\).
\eeq
The ladder expansion of $L^{(1,2)}$ at LL$x$ then has the form
\begin{align}
\label{eq:sigma3}
&\sigma^{n,m}\(N,\frac{\muf^2}{Q^2},\as;\epsilon\)=\int_0^\infty
\left[\gamma\(N,\(\frac{\muf^2}{Q^2\xi_n}\)^\epsilon,\as;\epsilon\)\right]\frac{d\xi_n}{\xi_n^{1+\epsilon}}
\times
\notag\\ 
&\times \int_0^\infty
\left[\gamma\(N,\(\frac{\muf^2}{Q^2\bar{\xi}_m}\)^\epsilon,\as;\epsilon\)\right]\frac{d\bar{\xi}_m}{\bar{\xi}_m^{1+\epsilon}}
C\(N,\xi,\bar{\xi},\as; \epsilon\)\times\\ 
&\times\int_0^{\xi_n}\left[\gamma\(N,\(\frac{\muf^2}{Q^2\xi_{n-1}}\)^\epsilon,\as;\epsilon\)\right]\frac{d\xi_{n-1}}{\xi_{n-1}^{1+\epsilon}}\times
\dots \times
\int_0^{\xi_2}\left[\gamma\(N,\(\frac{\muf^2}{Q^2\xi_1}\)^\epsilon,\as;\epsilon\)\right]\frac{d\xi_1}{\xi_1^{1+\epsilon}}\times\notag\\ 
&\times \int_0^{\bar{\xi}_m}
\left[\gamma\(N,\(\frac{\muf^2}{Q^2\bar{\xi}_{m-1}}\)^\epsilon,\as;\epsilon\)\right]\frac{d\bar{\xi}_{m-1}}{\bar{\xi}_{m-1}^{1+\epsilon}}\times
\dots \times \int_0^{\bar{\xi}_2}
\left[\gamma\(N,\(\frac{\muf^2}{Q^2\bar{\xi}_1}\)^\epsilon,\as;\epsilon\)\right]\frac{d\bar{\xi}_1}{\bar{\xi}_1^{1+\epsilon}}.\notag
\end{align}

Factorization is performed by requiring Eq.~\eqref{eq:sigma3} to be
finite after each $\xi_i$ or $\bar{\xi}_j$ integration. This leave a
single $n+m$-th order $\epsilon$ pole in the cross-section that can be
subtracted using the standard \MSbar~prescription (see Appendix A of
Ref.~\cite{Caola:2010kv}). After iterative subtraction of the first
$n-1$ and $m-1$ singularities we get 
\begin{align}
\label{eq:sigmanmsubtracted}
&\sigma^{n,m}\(N,\frac{\muf^2}{Q^2},\as;\epsilon\)= \left[\gamma\(N,\(\frac{\muf^2}{Q^2}\)^\epsilon,\as;\epsilon\)\right]^2\int_0^\infty \frac{d\xi_n}{\xi_n^{1+\epsilon}} \int_0^\infty \frac{d\bar{\xi}_m}{\bar{\xi}_m^{1+\epsilon}} C\(N,\xi_n,\bar{\xi}_m,\as; \epsilon\)\times\notag\\
&\times \frac{1}{\(n-1\)!}\frac{1}{\epsilon^{n-1}}\left[\sum_j\frac{\tilde{\gamma}_j\(N,\as;0\)}{j}\(1-\(\frac{\muf^2}{Q^2\xi_n}\)^{j\epsilon}\frac{\tilde{\gamma}_i\(N,\as;\epsilon\)}{\tilde{\gamma}_i\(N,\as;0\)}\)\right]^{n-1}\times \\
&\times \frac{1}{\(m-1\)!}\frac{1}{\epsilon^{m-1}}\left[\sum_l\frac{\tilde{\gamma}_l\(N,\as;0\)}{l}\(1-\(\frac{\muf^2}{Q^2\bar{\xi}_m}\)^{l\epsilon}\frac{\tilde{\gamma}_i\(N,\as;\epsilon\)}{\tilde{\gamma}_l\(N,\as;0\)}\)\right]^{m-1}.\notag
\end{align}
where we have introduced  the expansion
\beq
\label{eq:tildegamma}
\gamma\(N,\(\frac{\muf^2}{Q^2\xi}\)^\epsilon,\as;\epsilon\)=\sum_{j=0}^\infty \tilde{\gamma}_j\(N,\as;\epsilon\)\(\frac{\muf^2}{Q^2\xi}\)^{j\epsilon}.
\eeq

Summing over $n$ and $m$ the collinear singularities exponentiate:
\begin{align}
\label{eq:sigmares1}
\sigma_{\rm res}&=\sum_{n,m=0}^{\infty}\sigma^{n,m}=\gamma\(N,\(\frac{\muf^2}{Q^2}\)^\epsilon,\as;\epsilon\)^2\int_0^\infty\frac{d\xi}{\xi^{1+\epsilon}}\int_0^\infty\frac{d\bar{\xi}}{\bar{\xi}^{1+\epsilon}}\,C\(N,\xi,\bar{\xi},\as;\epsilon\)\times \notag\\
&\times \exp\left[\frac{1}{\epsilon}\sum_j\frac{\tilde{\gamma}_j\(N,\as;0\)}{j}\(1-\(\frac{\muf^2}{Q^2\xi}\)^{j\epsilon}\frac{\tilde{\gamma}_j\(N,\as;\epsilon\)}{\tilde{\gamma}_j\(N,\as;0\)}\)\right]\times \\
&\times \exp\left[\frac{1}{\epsilon}\sum_l\frac{\tilde{\gamma}_l\(N,\as;0\)}{l}\(1-\(\frac{\muf^2}{Q^2\bar{\xi}}\)^{l\epsilon}\frac{\tilde{\gamma}_l\(N,\as;\epsilon\)}{\tilde{\gamma}_l\(N,\as;0\)}\)\right].\notag
\end{align}
The limit  $\epsilon\to 0$ can then be taken after expanding
\beq
\tilde{\gamma}_i \equiv \tilde{\gamma}_i\(N,\as\)+\epsilon\dot{\tilde{\gamma}}_i\(N,\as\)+\epsilon^2\ddot{\tilde{\gamma}}_i\(N,\as\)+\dots,
\eeq
with the result
\begin{align}
\label{eq:sigmares2}
\sigma_{\rm res}\(N,\as\)=\gamma\(N,\as\)^2\mathcal{R}\(N,\as\)^2&\int_0^\infty d\xi\,\xi^{\gamma\(N,\as\)-1}\int_0^{\infty}d\bar{\xi}\,\bar{\xi}^{\gamma\(N,\as\)-1}\, C\(N,\xi,\bar{\xi},\as\)\notag\\
&\times\exp\left[2\gamma\(N,\as\)\ln\frac{Q^2}{\muf^2}\right]
\end{align}
with~\cite{Caola:2010kv}
\beq
\mathcal{R}\(N,\as\)\equiv\exp\left[-\sum_i \frac{\dot{\tilde{\gamma}}_i\(N,\as\)}{i}\right].
\eeq

Equation~\eqref{eq:sigmares2} is the resummed form of the partonic
cross section at LL$x$ in the \MSbar~scheme, after factorization of all
singularities. The factor $\mathcal{R}$ depends on the choice of
factorization scheme~\cite{Caola:2010kv,Catani:1990eg}; further
scheme changes may be performed by redefining the parton distribution
of the gluon by a generic LL$x$ function
$\mathcal{N}\(N,\as\)$~\cite{Ball:1995np}, after which all the scheme 
dependence can be collected in a prefactor
\beq
R\(N,\as\)=\mathcal{R}\(N,\as\)\mathcal{N}\(N,\as\).
\eeq
Choosing  $\muf^2=Q^2$ the final form of the resummed inclusive cross section is
\beq
\label{eq:sigmaresfinal}
\sigma_{\rm res}\(N,\as\)=\gamma\(\frac{\as}{N}\)^2 R\(\frac{\as}{N}\)^2\int_0^\infty d\xi\,\xi^{\gamma\(\frac{\as}{N}\)-1}\int_0^{\infty}d\bar{\xi}\,\bar{\xi}^{\gamma\(\frac{\as}{N}\)-1}\, C\(N,\xi,\bar{\xi},\as\).
\eeq
where we have explicitly indicated that, at LL$x$, $\gamma$ and $R$ only
depend on  the ratio $\as/N$.
 
In order to make contact with the approach of
Ref.~\cite{Catani:1990eg}, it is useful to rewrite the resummed cross-section
  in terms of the so-called impact factor,
defined as
\beq
h\(N,M_1,M_2,\as\)=M_1 M_2\,R\(M_1\)R\(M_2\)\int_0^\infty d\xi\,\xi^{M_1-1}\int_0^\infty d\bar{\xi}\,\bar{\xi}^{M_2-1}\,C\(N,\xi,\bar{\xi},\as\),
\eeq
in terms of which the cross-section Eq.~\eqref{eq:sigmaresfinal} has
the form
\beq
\sigma_{\rm res}\(N,\as\)=h\(N,\gamma\(\frac{\as}{N}\),\gamma\(\frac{\as}{N}\),\as\).
\eeq
The explicit expressions of the LL$x$ anomalous dimension $\gamma$ and
the factorization-scheme dependent function $R$ can be found e.g. in
Ref.~\cite{Catani:1994sq}.


\section{The transverse momentum distribution}
\label{sec:HFTM}
Having briefly reviewed the approach of Ref.~\cite{Caola:2010kv} to
high-energy resummation, we now extend it to transverse momentum
distributions: the generalization turns out to be in fact completely
straightforward, once the kinematics is properly understood.

 We consider again the process Eq.~\eqref{eq:process}, but now
assuming that $\mathcal{S}$ has fixed transverse momentum $\pt$.
Clearly (see  Fig.~\ref{fig:hardladder2})  $\pt$ is the sum of the
transverse momenta  $\kt$ and $\barkt$, of the gluons which connect
the hard part to the ladder, so in the high-energy limit Eq.~\eqref{eq:helim}
it must satisfy
\beq
\label{eq:ptconstraint}
\frac{\pt^2}{s} \ll 1.
\eeq
The  factorization
Eq.~\eqref{eq:sigmahighenergy}, which was derived by power counting
from the conditions  Eq.~\eqref{eq:helim} still holds, but now with
a kinematic constraint relating  $\pt$ to $\kt$ and $\barkt$:
\begin{align}
\label{eq:ptsigmahighenergy}
\frac{d\sigma}{d\pt^2}=Q^2\,\int &\left[\frac{x}{2 z \bar{z}} H_{\parallel,\parallel}\(\frac{x}{z\bar{z}},\frac{\kt^2}{Q^2},\frac{\barkt^2}{Q^2},\Omega_\mathcal{S},\as\)\right]\delta\(\pt^2-\kt^2-\barkt^2-2\sqrt{\kt^2\barkt^2}\cos\theta\)\notag\\
&\left[2\pi L^{(1)}_{\parallel}\(\frac{\muf^2}{\kt^2},\as\)\right]\left[2\pi L^{(2)}_{\parallel}\(\frac{\muf^2}{\barkt^2},\as\)\right]\frac{dz}{z}\frac{d\bar{z}}{\bar{z}}\frac{d\kt^2}{\kt^2}\frac{d\barkt^2}{\barkt^2}\frac{d\theta}{2\pi}\frac{d\bar{\theta}}{2\pi}.
\end{align}
The constraint is a simple momentum conservation delta as a
consequence of the fact that radiation is entirely contained in the
ladders, and it does not take place from the hard part; without loss
of generality, we have chosen
$\theta$ as the angle between the directions of $\mathbf{k}$ and
$\mathbf{\bar{k}}$. 

We then  define a  $\pt$-dependent coefficient function 
\begin{align}
\label{eq:Cpt}
&C_{\pt}\(\frac{x}{z \bar{z}},\xi,\bar{\xi},\xi_p,\as\)=\notag\\&\qquad=\int \frac{d\theta}{2\pi}\frac{d\bar{\theta}}{2\pi}\,\frac{x}{2 z \bar{z}}H_{\parallel,\parallel}\(\frac{x}{z \bar{z}},\xi,\bar{\xi},\Omega_\mathcal{S},\as\)\delta\(\xi_p-\xi-\bar{\xi}-2\sqrt{\xi\bar{\xi}}\cos\theta\)\notag\\
&\qquad=\int
\frac{d\theta}{2\pi}\frac{d\bar{\theta}}{2\pi}\,\frac{x}{2 z \bar{z}}
\left[\mathcal{P}^{\mu\nu}\mathcal{P}^{\bar{\mu}\bar{\nu}}H_{\mu\nu\bar{\mu}\bar{\nu}}\right]\delta\(\xi_p-\xi-\bar{\xi}-2\sqrt{\xi\bar{\xi}}\cos\theta\)
\end{align}
where we have introduced a further dimensionless variable
\begin{align}\label{eq:xipdef}
\xi_p=\frac{\pt^2}{Q^2},
\end{align}
on top of  $\xi$, $\bar \xi$ Eq.~\eqref{eq:xidef}.
In terms of $C_{\pt}$, Eq.~\eqref{eq:Cpt} becomes
\begin{align}
\frac{d\sigma}{d\xi_p}=\int &C_{\pt}\(\frac{x}{z \bar{z}},\xi,\bar{\xi},\bar{\xi_p},\as\)\notag\\
&\times\left[2\pi L^{(1)}_{\parallel}\(\frac{\muf^2}{Q^2 \xi},\as\)\right]\left[2\pi L^{(2)}_{\parallel}\(\frac{\muf^2}{Q^2 \bar{\xi}},\as\)\right]\frac{dz}{z}\frac{d\bar{z}}{\bar{z}}\frac{d\xi}{\xi}\frac{d\bar{\xi}}{\bar{\xi}}.
\end{align}
The coefficient function $C_{\pt}$ is 
the transverse momentum distribution for the production of
$\mathcal{S}$ from two off-shell gluons with transverse momenta $\mathbf{k}$ and $\mathbf{\bar{k}}$.

\begin{figure}[htb]
\centering
\label{fig:generalizedladderpt}
\includegraphics[width=0.8\textwidth]{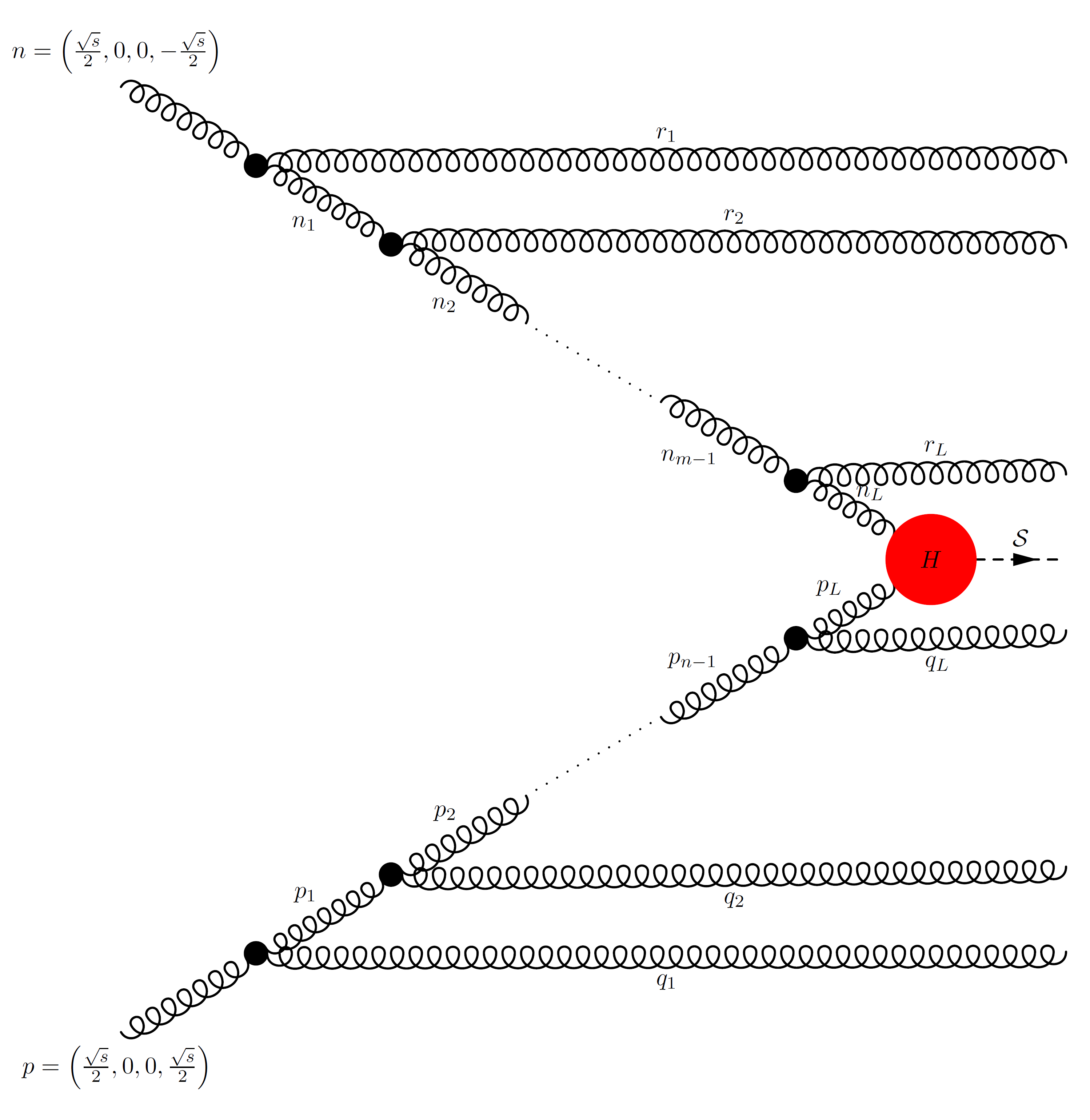}
\caption{Kinematics of the ladder. The blob at each emission vertex
  denotes inclusion of LL$x$ $s$- and $t$-channel gluon radiation to
  all orders.}
\end{figure}

We now turn to the ladders. Each insertion of the LL$x$ kernel $K$
Eq.~(\ref{eq:llxk}) includes an infinite series of $s$- and
$t$-channel branchings~\cite{Ball:1997vf}, which can be viewed as a
single effective emission vertex.
The momenta of the gluons
$q_1$,\dots,$q_L$ and $r_1$,\dots,$r_L$ respectively 
radiated from each of the two
rails of the ladder, and of the gluons $p_1$,\dots,$p_L$ and
$n_1$,\dots,$n_L$ respectively propagating along them, in the Sudakov
parametrization in the high-energy limit  can be written  
as (see Fig.~\ref{fig:generalizedladderpt})
\begin{subequations}
\label{eq:kinematics}
\begin{align}
p_1&=z_1 p - \mathbf{k}_1\\
q_1&=\(1-z_1\) p + \mathbf{k}_1\\
p_2&=z_2 z_1 p - \mathbf{k}_2 \\
q_2&=\(1-z_1z_2\)z_1 p + \mathbf{k}_2-\mathbf{k}_1\label{eq:q2kin}\\
&\dots\dots\dots\\
p_L&=z p - \mathbf{k} \\
q_L&=\(1-z\)p +\mathbf{k}-\mathbf{k_{n-1}}\\
\quad\notag\\
n_1&=\bar{z}_1 p - \mathbf{\bar{k}}_1\\
r_1&=\(1-\bar{z}_1\) p + \mathbf{\bar{k}}_1\\
n_2&=\bar{z}_2 \bar{z}_1 p - \mathbf{\bar{k}}_2 \\
r_2&=\(1-\bar{z}_1\bar{z}_2\)\bar{z}_1 p +\mathbf{\bar{k}}_2-\mathbf{\bar{k}}_1\\
&\dots\dots\dots\\
n_L&=\bar{z} n - \mathbf{\bar{k}} \\
r_L&=\(1-\bar{z}\)n +\mathbf{\bar{k}-\mathbf{\bar{k}_{m-1}}}.
\end{align}
\end{subequations}

The crucial observation here is that 
the momenta of the emitted gluons $q_1$ and $r_i$ are integrated
over. So, for instance, the transverse momentum of the second emitted
gluon is an integration variable, and we can equivalently choose it as
$\mathbf{k}_2$ or, shifting the integration variable, as 
$ \mathbf{k}_2-\mathbf{k}_1$, as in Eq.~(\ref{eq:q2kin}). With the
choice of integration variables of Eq.~\eqref{eq:kinematics}, it is
manifest that
all
the transverse momenta $\mathbf{k}_i$ and $\mathbf{\bar{k}}_j$ are
independent, with the only ordering constraints 
$k_{\sss T,1}^2 \ll k_{\sss T,2}^2 \ll \dots \ll k_{\sss T}^2$ and
 $\bar{k}_{\sss T,1}^2 \ll \bar{k}_{\sss T,2}^2 \ll \dots \ll
\bar{k}_{\sss T}^2$. The fixed value of $\pt^2$ of the final state
$\mathcal S$ thus only constrains 
the transverse components of the momenta
$p_L$
 $n_L$ of the two gluons entering the hard part $H$. The dependence on
the longitudinal momentum fractions in  Eq.~\eqref{eq:kinematics} is
immaterial for our purposes, and was discussed in Ref.~\cite{Caola:2010kv}.

It follows that we can compute the ladders as in the inclusive case, the
only difference being in the integration over the transverse momenta
of the two gluons connecting each ladder to the hard part:
we iterate the
kernel $K$ and sum over all possible
insertions.
The regularized contribution to the transverse momentum distribution
when the kernel $K$ is inserted $n$-th times on one leg and $m$-th
times on the other leg, after the iterative subtraction of the first
$n-1$ and $m-1$ collinear singularities has the form
\begin{align}
\label{eq:pt sigmanmsubtracted}
&\frac{d\sigma^{n,m}}{d\xi_p}\(N,\frac{\muf^2}{Q^2},\xi_p,\as;\epsilon\)= \left[\gamma\(N,\(\frac{\muf^2}{Q^2}\)^\epsilon,\as;\epsilon\)\right]^2\notag\\
&\times \int_0^\infty \frac{d\xi_n}{\xi_n^{1+\epsilon}} \int_0^\infty \frac{d\bar{\xi}_m}{\bar{\xi}_m^{1+\epsilon}} C_{\pt}\(N,\xi_n,\bar{\xi}_m,\xi_p,\as; \epsilon\)\times\notag\\
&\times
\frac{1}{\(n-1\)!}\frac{1}{\epsilon^{n-1}}\left[\sum_j\frac{\tilde{\gamma
     }_j\(N,\as;0\)}{j}\(1-\(\frac{\muf^2}{Q^2\xi_n}\)^{j\epsilon}\frac{\tilde{\gamma}_i\(N,\as;\epsilon\)}{\tilde{\gamma}_i\(N,\as;0\)}\)\right]^{n-1}\times \notag\\
&\times \frac{1}{\(m-1\)!}\frac{1}{\epsilon^{m-1}}\left[\sum_l\frac{\tilde{\gamma}_l\(N,\as;0\)}{l}\(1-\(\frac{\muf^2}{Q^2\bar{\xi}_m}\)^{l\epsilon}\frac{\tilde{\gamma}_i\(N,\as;\epsilon\)}{\tilde{\gamma}_l\(N,\as;0\)}\)\right]^{m-1}.
\end{align} 
with $\tilde{\gamma}$ defined as in  Eq.~\eqref{eq:tildegamma}. 

Summing over emissions the result exponentiates, as in the inclusive
case;  the only nontrivial difference is the delta constraint
which is included in the $\pt$-dependent coefficient function
Eq.~\eqref{eq:Cpt}: 
\beq
\label{eq:ptsigmaresfinal}
\frac{d\sigma_{\rm res}}{d\xi_p}\(N,\xi_p,\as\)=\gamma\(\frac{\as}{N}\)^2 R\(\frac{\as}{N}\)^2\int_0^\infty d\xi\,\xi^{\gamma\(\frac{\as}{N}\)-1}\int_0^{\infty}d\bar{\xi}\,\bar{\xi}^{\gamma\(\frac{\as}{N}\)-1}\, C_{\pt}\(N,\xi,\bar{\xi},\xi_p,\as\).
\eeq 
Equation~\eqref{eq:ptsigmaresfinal} provides a resummed expression for
the transverse momentum distribution. Note that at LL$x$ if the
coefficient function is finite as $N\to0$ we can set $N=0$. While for total cross-sections this is not true
for pointlike interactions, we will show at the end of this section 
that this is always
true for transverse momentum distributions.

As in the inclusive case, this resummed result can be expressed
 in terms of an impact factor, now $\pt$-dependent:
\begin{align}
\label{eq:ptimpact}
h_{\pt}\(N,M_1,M_2,\xi_p,\as\)=&M_1 M_2 R\(M_1\)R\(M_2\)\notag\\
&\int_0^\infty d\xi\,\xi^{M_1-1}\int_0^\infty d\bar{\xi}\,\bar{\xi}^{M_2-1}\,C_{\pt}\(N,\xi,\bar{\xi},\xi_p,\as\)
\end{align}
by exploiting BFKL-DGLAP duality~\cite{Altarelli:1999vw} to set
\beq\label{eq:dual}
M_i=\gamma\(\frac{\as}{N}\)
\eeq
with the result
\beq
\label{eq:identification}
\frac{d\sigma_{res}}{d\xi_p}\(N,\xi_p,\as\)=h_{\pt}\(0,\gamma\(\frac{\as}{N}\),\gamma\(\frac{\as}{N}\),\xi_p,\as\),
\eeq
which is completely equivalent to the previous expression 
Eq.~\eqref{eq:ptsigmaresfinal}, having  explicitly set $N=0$.

We have thus come to the conclusion that high energy
resummation  of a transverse momentum distribution is 
obtained using the same formula as in the inclusive case, 
but with the total cross-section replaced by the corresponding
transverse-momentum distribution. This result is simple but 
powerful: in particular, it is worth noting that this means that the
nontrivial dependence on the transverse momentum is induced through
the kinematic constraint Eq.~\eqref{eq:Cpt} by the transverse momentum
of the incoming off-shell gluons, which in turn is determined through
Eqs.~(\ref{eq:ptimpact}-\ref{eq:identification}) by the LL$x$
anomalous dimension (i.e., equivalently, the BFKL kernel).

An immediate consequence of our derivation is  that the 
resummation of transverse
momentum distributions for lepto- or photo-production processes
reduces to that of the total cross-section. 
Indeed, when only one hadron is present in the initial
state Eq.~\eqref{eq:ptsigmaresfinal} reduces to
\beq
\label{eq:ptsigmalepto}
\frac{d\sigma^{\rm res}}{d\xi_p}\(N,\xi_p,\as\)=\gamma\(\frac{\as}{N}\)R\(\frac{\as}{N}\)\int_0^\infty d\xi\,\xi^{\gamma\(\frac{\as}{N}\)-1}C_{\pt}\(N,\xi,\xi_p,\as\),
\eeq
but in this case the momentum conservation constraint is trivial:
\beq
\label{eq:Cptlepto}
C_{\pt}\(N,\xi,\xi_p,\as\)=C\(N,\xi,\as\)\delta\(\xi_p-\xi\).
\eeq
so, substituting  Eq.~\eqref{eq:Cptlepto} in
Eq.~\eqref{eq:ptsigmalepto}, we
get the resummed result
\beq
\label{eq:ptsigmalepto2}
\frac{d\sigma^{\rm res}}{d\xi_p}\(N,\xi_p,\as\)=\gamma\(\frac{\as}{N}\)R\(\frac{\as}{N}\)\xi_p^{\gamma\(\frac{\as}{N}\)-1}C\(0,\xi_p,\as\),
\eeq
where the coefficient function $C$ is the same as in the inclusive case.

Finally, we consider quark-initiated hadro-production.  
As well-known~\cite{Catani:1994sq} the high energy
behaviour of quark channels can be deduced from that of the purely gluonic
channel by using the color-charge relation
$\gamma_{qg}=\frac{\Cf}{\Ca}\gamma_{gg}$, which holds at LL$x$ to all
orders in $\alpha_s$, and noting that
$\gamma_{gq}$ and $\gamma_{qq}$ are NLL$x$. It follows that
at LL$x$ a quark may turn into a gluon but a gluon cannot turn into a
quark. 
Hence, the computation of the resummed cross-section proceeds as for the
gluon channels, but with the subtraction of 
the contribution from diagrams where no emission takes place from the
quark leg,  since they are subleading in the high energy
regime~\cite{Catani:1994sq}. This leads to the following expressions
for the resummed  transverse-momentum  distributions in
quark-initiated channels:
\begin{subequations}
\label{eq:otherchannel}
\begin{align}
\(\frac{d\sigma^{\rm res}}{d\xi_p}\)_{gq}=&\frac{\Cf}{\Ca}\left[h_{\pt}\(0,\gamma\(\frac{\as}{N}\),\gamma\(\frac{\as}{N}\),\xi_p,\as\)-h_{\pt}\(0,\gamma\(\frac{\as}{N}\),0,\xi_p,\as\)\right],\\
\(\frac{d\sigma^{\rm res}}{d\xi_p}\)_{q\bar{q}}=&\(\frac{\Cf}{\Ca}\)^2\left[h_{\pt}\(0,\gamma\(\frac{\as}{N}\),\gamma\(\frac{\as}{N}\),\xi_p,\as\)-2h_{\pt}\(0,\gamma\(\frac{\as}{N}\),0,\xi_p,\as\)\right],
\end{align}
\end{subequations}
where $h_{\pt}$ is the gluon-channel impact factor
Eq.~\eqref{eq:ptimpact},  and the color-charge factor
$\frac{\Cf}{\Ca}$ is due to the presence of $\gamma_{qg}$ in the first
gluon emission. 

The total resummed cross-section can be obtained in each case by
integration of the transverse momentum distributions. The high energy
behaviour of the total cross-section, as
well-known~\cite{Catani:1990xk, Catani:1990eg}, is single-logarithmic,
or double logarithmic, according to whether the hard interaction is
pointlike or not:
\begin{numcases}{\sigma\apprlim{x\to0}\sigma_{\rm LO}\times}  \label{eq:incbehpt}
\delta\(1-x\)+\sum_{k=1}^\infty c_k\as^k\ln^{2k-1}\frac{1}{x}, & pointlike\\
\label{eq:incbehres}
\delta\(1-x\)+\sum_{k=1}^\infty d_k\as^k\ln^{k-1}\frac{1}{x}, &  resolved.
\end{numcases}
An interaction is pointlike if it does not resolve the $\pt$
dependence, i.e. more formally if it can be represented by the
insertion of a single local operator: in such case, the hard part is
independent of $\pt$, i.e. of $\xi_p$. All the $\xi_p$ dependence then
comes from the prefactors $\xi^\gamma$ in Eq.~\eqref{eq:ptimpact},
which are due to collinear radiation in the ladders: the
transverse momentum integration over
gluon radiation is undamped at high scale, and its logarithmic
divergence is cut off by
center-of-mass energy. In Mellin space, this corresponds to the fact
that  the impact factor diverges as $N\to0$. In such case, 
expansion in powers of $\alpha_s$ leads to double poles in $N$ and
thus double logs in $x$.

The resummed transverse momentum distribution always displays single
 logarithmic behaviour, because the $\xi_p\to\infty$ limit is never
 reached. However, when the interaction is pontlike,
 the coefficients grow
 logarithmically with $\pt$ (or equivalently with $\xi_p$),
while  in the resolved (non-pointlike) case the coefficients
$d_k\(\xi_p\)$  as $\xi_p\to\infty$ vanish at least as a power of
$\xi_p^{-1}$ in such a way that the integral over all transverse
momenta is finite:
\begin{numcases}
{\frac{d\sigma}{d\xi_p}\apprlim{x\to0}\frac{\sigma_{LO}}{\xi_p}\times}
\label{eq:ptbehpt}
\sum_{k=1}^\infty \as^k\ln^{k-1}\frac{1}{x}\sum_{n=0}^{k-1}c_{kn}\ln^n\xi_p, &pointlike\\
\label{eq:ptbehptres}
\sum_{k=1}^\infty d_k\(\xi_p\)\as^k\ln^{k-1}\frac{1}{x}, &resolved.
\end{numcases}

\section{Higgs production in gluon fusion}
\label{sec:Higgs}
\begin{figure}[htb]
\label{fig:LOhiggs}
\centering
\includegraphics[width=1.0\textwidth]{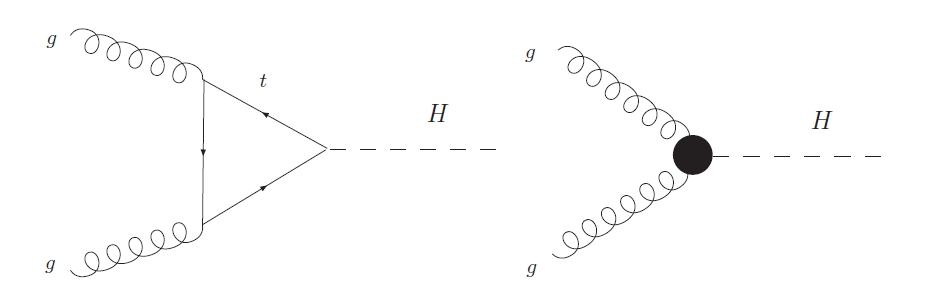}
\caption{Born Level diagrams for Higgs boson production in $gg$ channel, respectively in the full (left) and effective theory (right).}
\end{figure}

We now use the general result Eq.~\eqref{eq:identification}
and compute the high energy behaviour at LL$x$ of the
transverse momentum distribution for  Higgs production in gluon
fusion (see Fig.~\ref{fig:LOhiggs}). The full result is only known
at LO~\cite{Baur:1989cm}. However, in the effective field theory in
which the mass of the quark in the loop goes to infinity, it is
known in fully analytic form up to
NLO~\cite{Glosser:2002gm,Ravindran:2002dc}, and at NNLO with a numerical
evaluation of the phase space integrals~\cite{Boughezal:2015dra}. 

Here we will only consider the case of the effective field
theory: we  first determine the full resummed result, and then we
expand it out up to $\Ord(\alpha_s^4)$. This illustrates our
methodology, it provides a nontrivial check of it, and yields  a
prediction for the next fixed order. 

As explained in the previous section, in order to determine the
$\pt$-dependent impact factor $h_{\pt}$, Eq.~\eqref{eq:ptimpact}, we
must determine the transverse momentum distribution $C_{\pt}$ for the process
\beq
g^*\(p_L\) + g^*\(n_L\) \to H\(p_\mathcal{S}\)
\eeq
with incoming off-shell gluons.
The color-averaged squared matrix element in the effective theory is~\cite{Hautmann:2002tu,Marzani:tesi}
\beq
\abs{\mathcal{M}}^2=\frac{\as^2}{32}\frac{\sqrt{2}\Gf}{9\pi^2}\frac{\(\mathbf{k}\cdot \mathbf{\bar{k}}\)^2}{\abs{\mathbf{k}}^2\abs{\mathbf{\bar{k}}^2}}\(\frac{\mH^2}{\tau}\)^2
\eeq
where ${\bf k}$ and ${\bf\bar{k}}$ are respectively
the transverse components of $p_L$ and $n_L$ Eq.~\eqref{eq:sudakovparametrization}, and $\tau=\frac{x}{z\bar{z}}$.

The coefficient function $C_{\pt}$ is found by providing
necessary phase space factor, and performing a Mellin transform:
\begin{align}
\label{eq:Cpthiggs}
&C_{\pt}\(N,\xi,\bar{\xi},\xi_p,\as\)=\frac{\as^2 \sqrt{2}\Gf}{288\pi}\int_0^1 d\tau\,\tau^{N-1}\int_0^{2\pi} \frac{d\theta}{2\pi} \frac{\cos^2\theta}{\tau}\,\delta\(\frac{1}{\tau} - 1- \xi_p\)\notag\\
&\qquad\qquad\qquad\delta\(\xi_p-\xi-\bar{\xi}-2\sqrt{\xi\bar{\xi}}\cos\theta\)\nonumber\\
&\quad=2\sigma_{\sss \rm LO}\;\,\int_0^1d\tau\,\tau^{N-2}\,\delta\bigg(\frac{1}{\tau}-1-\xi_p\bigg)\int_0^{2\pi}\frac{d\theta}{2\pi}\cos^2\theta\,\delta\(\xi_p-\xi-\bar{\xi}-2\sqrt{\xi\bar{\xi}}\cos\theta\)
\end{align}
where $\xi$,  $\bar{\xi}$ and $\xi_p$ were defined in
Eqs.~(\ref{eq:xidef}-\ref{eq:xipdef})
and
\beq
\sigma_{\sss \rm LO}=\frac{\as^2\sqrt{2}\Gf}{576\pi}
\eeq
is the leading-order inclusive cross-section.

The integrals in $\tau$ and $\theta$ in Eq.~\eqref{eq:Cpthiggs} can
be performed explicitly, with the result
\begin{align}
\label{eq:hpthiggs}
&h_{\pt}\(N,M_1,M_2,\xi_p,\as\)=\frac{\sigma_{\sss \rm LO}}{2\pi\(1+\xi_p\)^N}M_1 M_2 R\(M_1\)R\(M_2\)\notag\\
&\qquad\qquad\int_0^\infty d\xi\,\xi^{M_1-2}\int_{\(\sqrt{\xi_p}-\sqrt{\xi}\)^2}^{\(\sqrt{\xi_p}+\sqrt{\xi}\)^2}d\bar{\xi}\,\bar{\xi}^{M_2-2}
\frac{\(\xi_p-\xi-\bar{\xi}\)^2}{\sqrt{2\bar{\xi}\xi+2\xi\xi_p+2\bar{\xi}\xi_p-\xi_p^2-\xi^2-\bar{\xi}^2}}.
\end{align}
Changing variables 
\begin{equation}
\xi=\xi_p\,\xi_1,\qquad  \bar{\xi}=\xi_p\,\xi_2,
\end{equation}
the dependence on 
$\xi_p$ can be taken outside the integral in Eq.~\eqref{eq:hpthiggs}:
\begin{equation}
\label{eq:hpthiggs2}
h_{\pt}\(N,M_1,M_2,\xi_p,\as\)=\sigma_{\sss \rm LO}\frac{\xi_p^{M_1+M_2-1}}{\(1+\xi_p\)^N}I\(M_1,M_2\),
\end{equation}
and the integral
\begin{align}
\label{eq:I}
&I\(M_1,M_2\)=M_1 M_2\, R\(M_1\)R\(M_2\)\\
&\qquad \int_0^\infty d\xi_1\,\xi_1^{M_1-2}\,\int_{\(1-\sqrt{\xi_1}\)^2}^{\(1+\sqrt{\xi_1}\)^2}d\xi_2\,\xi_2^{M_2-2}\frac{\(1-\xi_1-\xi_2\)^2}{\sqrt{2\xi_1\xi_2+2\xi_1+2\xi_2-1-\xi_1^2-\xi_2^2}}
\end{align}
does not depend on $\xi_p$.

The integrals over $\xi_1$ and $\xi_2$ in $I$ are computed in 
Appendix~\ref{app:Higgsimpact}; substituting the result
[Eq.~\eqref{eq:ires}] in Eq.~\eqref{eq:hpthiggs2} we finally find that
the impact factor is given by
\begin{align}
\label{eq:hpthiggsfinal}
&h_{\pt}\(N,M_1,M_2,\xi_p,\as\)=R\(M_1\)R\(M_2\)\sigma_{\sss \rm LO}\frac{\xi_p^{M_1+M_2-1}}{\(1+\xi_p\)^N}\notag\\
&\qquad\left[\frac{\Gamma\(1+M_1\)\Gamma\(1+M_2\)\Gamma\(2-M_1-M_2\)}{\Gamma\(2-M_1\)\Gamma\(2-M_2\)\Gamma\(M_1+M_2\)}\(1+\frac{2M_1M_2}{1-M_1-M_2}\)\right].
\end{align}
The fact that the $\xi_p$ dependence is entirely contained in a
prefactor is  a consequence of the fact that in the effective theory
the interaction is pointlike, and thus the transverse momentum
dependence is entirely due  to collinear radiation, as discussed in
the end of Sect.~\ref{sec:HFTM}. 
The resummed result is found from Eq.~\eqref{eq:hpthiggsfinal} by letting $N=0$ and by
substituting for $M_i$ the LL$x$ anomalous dimension 
Eq.~\eqref{eq:dual}, according to Eq.~\eqref{eq:ptimpact}. 
Our result manifestly reproduces the expected all-order
behaviour Eq.~\eqref{eq:ptbehpt}. 

We may check that integration of the transverse momentum dependent
impact factor reproduces the known inclusive result: using the integral
\beq
\label{eq:int}
\int_0^\infty d\xi_p \frac{\xi_p^{M_1+M_2-1}}{\(1+\xi_p\)^N}=\frac{\Gamma\(M_1+M_2\)\Gamma\(N-M_1-M_2\)}{\Gamma\(N\)}
\eeq
in Eq.~\eqref{eq:hpthiggsfinal} we obtain the inclusive  impact factor 
as given in Eq.~(5.33) of Ref.~\cite{Marzani:tesi}.

We can now expand our result in powers of $\alpha_s$ in order to
compare to known fixed-order expressions. 
We get
\beq
\label{eq:sigmaresx}
\frac{d\sigma}{d\xi_p}\(N,\as\)=\sigma_{LO}\sum_{k=1}^\infty
C_k\(\xi_p\)\as^k\frac{\ln^{k-1} x}{k-1!}
\eeq
with
\begin{subequations}
\label{eq:Ci}
\begin{align}
C_1\(\xi_p\)=&\frac{2\Ca}{\pi}\frac{1}{\xi_p}\label{eq:Ci0}\\
C_2\(\xi_p\)=&\frac{4\Ca^2}{\pi^2}\frac{\ln\xi_p}{\xi_p}\label{eq:Ci1}\\
C_3\(\xi_p\)=&\frac{2\Ca^3}{\pi^3}\frac{1+2\ln^2\xi_p}{\xi_p}\label{eq:Ci2}\\
C_4\(\xi_p\)=&\frac{4\Ca^4}{\pi^4}\frac{3+3\ln\xi_p+2\ln^3\xi_p+17\zeta_3}{3\xi_p}.
\end{align}
\end{subequations}
Equation~\eqref{eq:sigmaresx} gives the result in the gluon channel;
results in channels involving quarks can be obtained using
Eq.~\eqref{eq:otherchannel}.

\begin{figure}[hb]
\centering
\includegraphics[width=0.48\textwidth]{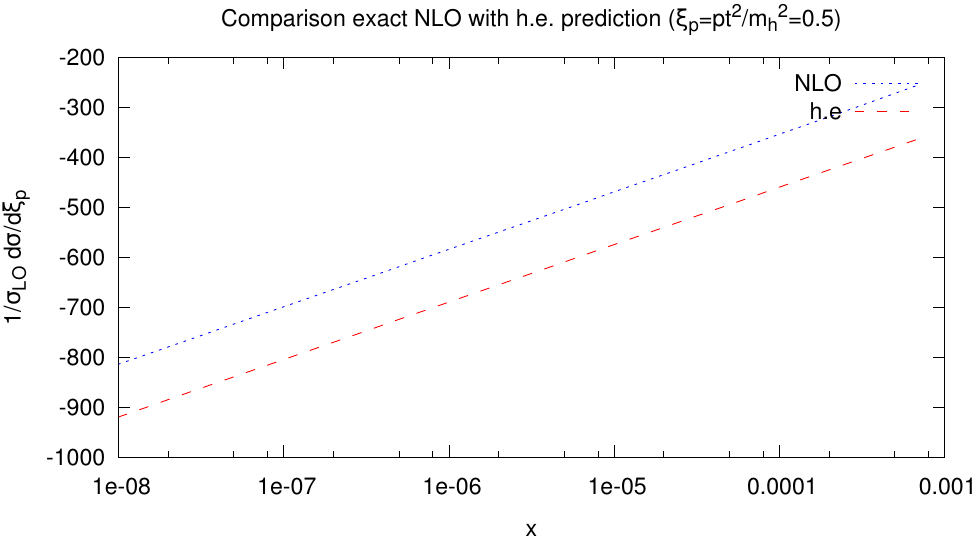}
\includegraphics[width=0.48\textwidth]{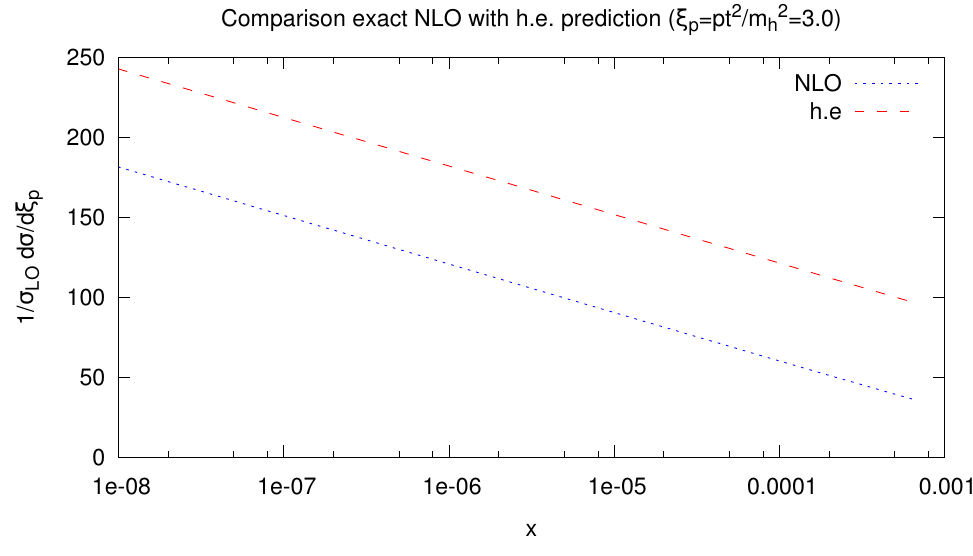}
\caption{NLO contribution to the transverse momentum distribution for
  Higgs production in gluon fusion, normalized to $\sigma_{LO}$,
  compared to the  high energy prediction
  $C_2\(\xi_p\)$ Eq.~\eqref{eq:Ci1} for two different fixed values of
  $\xi_p=0.5$ and $\xi_p=3.0$ .}\label{fig:NLOgg}
\end{figure}
Comparison to the LO exact result can be performed analytically.
The LO double-differential transverse momentum and rapidity
distribution in the effective field theory in the gluon-gluon channel
is given by~\cite{Ellis:1987xu,Baur:1989cm} 
\beq
  \frac{d\sigma^{(0)}}{d\xi_p dy}\(x,\xi_p,y\)=\sigma_{LO}	\frac{\as\Ca}{2\pi}x\frac{x^4+1+\(\frac{t}{s}\)^4+\(\frac{u}{s}	\)^4}{\frac{ut}{s^2}}\delta\(1+\frac{t}{s}+\frac{u}{s}-x\),
\eeq
where
\begin{align}
	x=&\frac{\mH^2}{s}\\
	\frac{t}{s}=&x-\sqrt{x}\sqrt{1+\xi_p}e^y \\
	\frac{u}{s}=&x+\sqrt{x}\sqrt{1+\xi_p}e^{-y}.
\end{align}
Integrating over rapidity we get
\beq
\frac{d\sigma^{(0)}}{d\xi_p}\(x,\xi_p,\as\)=\alpha_s\sigma_{LO}\frac{2\Ca}{\pi}\frac{1}{\xi_p}+\Ord\(x\)
\eeq
in agreement with Eq.~\eqref{eq:sigmaresx}.

At  NLO we compare to the full result numerically. The
 lengthy analytic expression for the double differential
distribution is given in Ref.~\cite{Glosser:2002gm}. 
We have integrated this numerically over 
rapidity $y$, retaining the full $x$ dependence: this is necessary
because, as discussed in
Ref.~\cite{Caola:2010kv}, terms which appear to be power-suppressed in
$x$ at the level of rapidity distribution lead to LL$x$
contributions upon integration.

\begin{figure}[htb]
\centering
\includegraphics[width=0.48\textwidth]{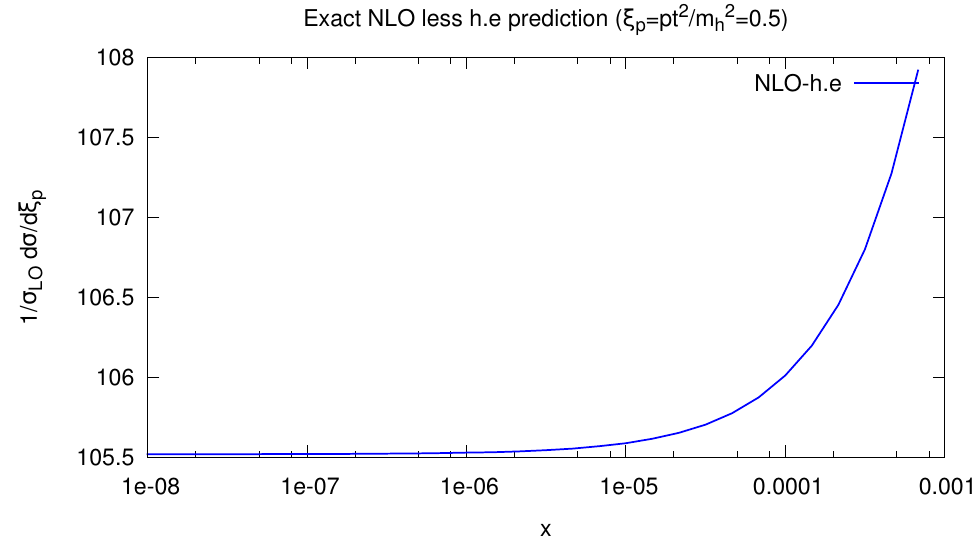}
\includegraphics[width=0.48\textwidth]{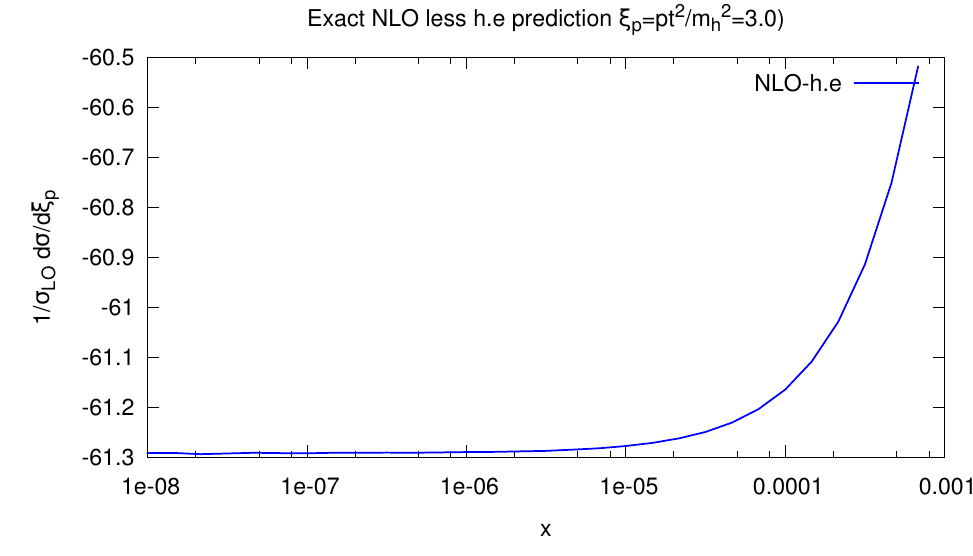}
\caption{Difference between the NLO fixed order result
  and the high energy prediction $C_2\(\xi_p\)$ Eq.~\eqref{eq:Ci1}
  shown in Fig.~\ref{fig:NLOgg}.}\label{fig:NLOdiff}
\end{figure}

The result of the integration  is plotted as a function of $\ln x$ in
Fig.~\ref{fig:NLOgg}, in the small $x$ region (blue line), together with
the  high-energy prediction Eq.~\eqref{eq:Ci1}, for two values of the
transverse momentum 
$\pt^2$. The difference between the two
curves is shown in Fig.~\ref{fig:NLOdiff}. It is clear that the
difference between the two curves tends to a constant as $x\to0$, thus proving 
perfect agreement between  the high-energy behaviour and the exact result.
We have repeated the comparison for a large number of values of
$\xi_p$, with the same result.
We have performed similar comparisons in the $gq$ and
$q\bar{q}$ channels, and find similarly good agreement.

A test at  NNLO is nontrivial due to the complexity of the 
exact result of Ref.~\cite{Boughezal:2015dra} which hampers its accurate
numerical evaluation in the high energy limit; it is very likely to
be possible thanks to the recent results of Ref.~\cite{Banfi:2015pju}.
However, the NNLO coefficient
can be tested by comparing to NNLL transverse momentum resummation, as
we discuss in the next section.

\section{Relation to transverse momentum resummation}
\label{sec:ptResum}

As we have argued on general grounds in Sect.~\ref{sec:HFTM},
Eq.~\eqref{eq:ptbehpt}, and seen explicitly in the case of Higgs in
gluon fusion in Sect.~\ref{sec:Higgs}, Eq.~\eqref{eq:hpthiggsfinal}
the high-energy transverse momentum distribution in the pointlike
limit displays an all-order single-logarithmic behaviour in
$\xi_p$. On the other hand, in the $\pt\to0$ limit (and not
necessarily at high energy) by standard Wilson
expansion arguments, the interaction can always be represented by a
local operator and the effect of any other scale (such as the heavy
quark masses) is entirely contained in a Wilson
coefficient. 

 Therefore, in the high energy limit, 
 the behaviour Eq.~\eqref{eq:ptbehpt} (seen in Eq.~\eqref{eq:hpthiggsfinal})
always holds when $\pt\to0$  i.e. when $\xi_p\to0$, even in the
resolved case,
 up to a prefactor (coming from the Wilson coefficient)
which in our LL$x$ limit is independent of $\alpha_s$ and only depends
on the scales which are integrated out in the effective field theory
(e.g., in the case discussed in the previous section on the ratios of the heavy quark
masses to the Higgs mass).

In this limit, however, hard cross sections are known to display
double logs  of the form $\frac{\ln^{2k-1} \xi_p}{\xi_p}$, which can be
resummed using now standard techniques~\cite{Collins:1984kg}: in
particular, N$^k$LL resummation allows one to predict the coefficients
of all contributions of the form $\as^n\,\frac{\ln^k\xi_p}{\xi_p}$
with $2(n-k)-1 \leq k \leq 2n-1$\footnote{Note
  that upon Fourier transformation, a $\frac{\ln^{k-1}\xi_p}{\xi_p}$ term
  corresponds to a $\ln^k b$ term, where  $b=|{\bf b}|$ is the
  impact parameter, see Eq.~(\ref{eq:qtres}).}
In the high energy limit, the hard cross section displays single logs
$\as^n\,\frac{\ln^n\xi_p}{\xi_p}$ Eq.~\eqref{eq:ptbehpt} . It
follows that at $\Ord(\alpha_s^n)$ the coefficient of the highest
power of $\ln \xi_p$ is predicted by N$^{n-1}$LL transverse
momentum resummation, with lower-order powers of $\ln \xi_p$ predicted
by increasingly subleading log resummation.
In particular,  the coefficients of $\ln^2\xi_p$ and $\ln\xi_p$ in $C_3$
Eq.~\eqref{eq:Ci2} are predicted by NNLL transverse momentum
resummation, thereby allowing us to also check this coefficient.

The LL$x$ result in the $\xi_p\to0$ limit, when taken to all orders in
$\alpha_s$, thus provides information on N$^k$LL transverse momentum resummation to all logarithmic orders
$0\le k\le\infty$ in  the $x\to0$ limit.
An immediate consequence of this is that the 
structure of transverse momentum resummation must be reproduced in the high-energy
limit.
This structure was fully elucidated only recently in
Refs.~\cite{Catani:2000vq,Catani:2011kr,Catani:2013tia}:
schematically, the contributions to the partonic cross section which
are singular as $\pt\to0$ have the form
\begin{align}\label{eq:qtres}
&\frac{d\sigma^{\rm res}_{a_1a_2}}{d{\bf \pt}}\(N,{\bf \pt},Q^2\)=\sum_{ij} 
\sigma^{(0)}_{ij}\int d^2{\bf b} e^{i {\bf b}\cdot{\bf \pt}}
S_{ij}(b^2,Q^2) \notag\\
&\quad\times  \sum_{lm}
  \left[H_{ij,lm}(\alpha_s) C_{il}(N,{\bf b})C_{jm}(N,{\bf
      b})+\bar{H}_{ij,lm}(\alpha_s) G_{il}(N,{\bf b})G_{jm}(N,{\bf
      b})\right]\times\\
&\qquad\times  \Gamma_{la_1}[\alpha_s,b^2,Q^2]\Gamma_{ma_2}[\alpha_s,b^2,Q^2],\notag
\end{align}
where $b=|{\bf b}|$; the sums over $(i,j)$ and $(l,m)$ run over parton
channels (quark and gluon), $\Gamma_i$ are standard QCD evolution
factors from scale $b$ to the hard scale $Q$ for the two incoming
partons $a_1$ $a_2$; $S_{ij}$ is a Sudakov evolution factor; and all
the process dependence is contained in the $N$-independent
hard functions $H$ and
$\bar H$ while the
partonic functions $C_{il}$ and $G_{il}$ on the two incoming legs  are universal. In the particular case
of quark-initiated channel, the $G_{il}$ functions vanish.

Equation~\eqref{eq:qtres} imposes on our resummed result the
nontrivial restriction that, in the  $\xi_p\to0$ limit, the dependence on
$M_1$, $M_2$ of the impact factor Eq.~\eqref{eq:ptimpact} 
factorized, in the sense that it can be written as a sum which reproduces
the schematic structure $C(M_1)C(M_2)+G(M_1)G(M_2)$  of the term in square
brackets of Eq.~(\ref{eq:qtres}).
This behaviour should hold in the small $\xi_p$
limit in general, and, for pointlike interactions, for all $\xi_p$.

Having understood the general structure of the constraints imposed by
the matching of transverse momentum resummation and high-energy resummation, we can
now check explicitly whether they are satisfied by our resummed
results.
In order to verify whether the structure Eq.~\eqref{eq:qtres} is
reproduced we must perform a Fourier transform of the resummed
cross-section. To this purpose, we define a $b$-space impact factor
\beq\label{eq:fourier}
h_{\pt}\(N,M_1,M_2, b,\as\)=\int_0^\infty d\xi_p\, 
J_0(\sqrt{\xi_p}\,b\,m_h)\,h_{\pt}\(N,M_1,M_2,\xi_p,\as\).
\eeq
The $b$-space
cross-section is obtained by performing the usual identification
Eq.~\eqref{eq:identification} with the  impact factor
Eq.~\eqref{eq:fourier}.

We get
\begin{align}\label{eq:bspaceres}
& h_{\pt}\(0,M_1,M_2, b,\as\)=\sigma_0
 e^{-(M_1+M_2)\ln\frac{  b^2 m_h^2}{4}}
R\(M_1\)R\(M_2\)  \notag\\&\qquad
\times\left[\frac{\Gamma\(1+M_1\)}{\Gamma\(1-M_1\)}
\frac{\Gamma\(1+M_2\)}{\Gamma\(1-M_2\)}+
M_1\frac{\Gamma\(1+M_1\)}{\Gamma\(2-M_1\)}
M_2\frac{\Gamma\(1+M_2\)}{\Gamma\(2-M_2\)}\right].
\end{align} We recognize the structure
Eq.~\eqref{eq:qtres}: the exponential prefactor corresponds to the
evolution factors $\Gamma_i$, as it is clear recalling that $M_i$ are
set equal to the anomalous dimensions while at LL$x$ level $\alpha_s$
does not run, and the term in square brackets reproduces the correct
structure of the universal partonic functions $C$ and $G$ of
Eq.~(\ref{eq:qtres}). Note that the hard function and the Sudakov
factor in Eq.~(\ref{eq:qtres}) do not depend on $N$; therefore, in the
high energy limit at LL$x$ only their trivial $O(\alpha_s^0)$ part
contributes. 

We thus see that indeed for pointlike interactions the structure of
the result Eq.~(\ref{eq:bspaceres}), as determined by transverse
momentum resummation, hold in fact for all $\xi_p$ and not just in the
small $\pt$ limit. On the other hand, we expect that in the
small $\pt$ limit the result found in the
full theory with exact top mass dependence will also reduce to the form
Eq.~(\ref{eq:bspaceres}). 

Having verified that our result has the correct structure fixed by
transverse momentum resummation constraints, we can check explicitly
the coefficients $C_i$ Eqs.~\eqref{eq:Ci}. Using the explicit
expression of NNLL resummation for Higgs
production~\cite{Bozzi:2005wk} in the small $x$ limit we get
\beq
\label{eq:sigmaptresum2}
\frac{d\sigma}{d\xi_p}\(N,\xi_p,\as\)=\sigma_{LO}\(1+\as^2\frac{\Ca^2}{N^2}\)\int_0^\infty db\,\frac{b}{2}\,J_0\(\sqrt{\xi_p}\,b\,m_h\)\exp\left[G^{h.e.}\(N,L\)\right]
\eeq
with
\beq
G^{h.e.}\(N,L\)=\frac{2\Ca}{N}\frac{\as}{\pi} L ,
\eeq
where 
\beq
\label{eq:L}
L \equiv L(b)=\ln b^2 m_h^2.
\eeq
Expanding the exponential and performing the Fourier transform in
Eq.~\eqref{eq:sigmaptresum2}   we immediately reproduce the
coefficients $C_1$, $C_2$, and 
the logarithmic contribution to $C_3$.
We have explicitly checked that the same holds in quark channels. We
conclude that our result is consistent with known results from
transverse momentum resummation.
 
\section{Outlook}
\label{sec:conclusion}
We have shown that transverse momentum distributions can be resummed
in the high energy limit in the same way as total cross-sections and
rapidity distributions, namely, by computing the corresponding
Born-level cross-section, but with incoming off-shell gluons.
The extra complexity due to the transverse momentum dependence is
entirely contained in the kinematic constraints which relates the
transverse momentum of the final state to the off-shellness of the
initial state, which is in turn re-expressed through high-energy
factorization in terms of the so-called BFKL, or LL$x$ anomalous
dimension. 
 
Because of its relative simplicity, our result provides a powerful
tool to obtain high-order information on collider processes. As a
first demonstration we have considered here the case of Higgs
production in gluon fusion in the pointlike limit. This is an
interesting case both for validation and conceptual reasons, because
full results are available to rather high perturbative orders, and
also because the pointlike limit, though displaying unphysical double
log behaviour at high energy, has a transverse momentum dependence
which can be related to that which is revealed in small transverse
momentum resummation. 

On the other hand, matching high energy  
to transverse momentum resummation, both
in the pointlike case and for the full theory, raises the interesting
question of combining the  two
resummations~\cite{marzaniprep}. However, it should be kept in mind
that for accurate phenomenology resummed results would have to be
combined with the running coupling resummation at high energy
discussed in  Refs.~\cite{Ball:2007ra,Altarelli:2008aj}.

On the other hand, the application of our technique to Higgs
production in gluon fusion when the full dependence on the top mass is
retained appears to be especially interesting
as a way to
gain information on higher order
corrections. Indeed, only the leading order result is known in this
case, while  the pointlike
approximation is known~\cite{DelDuca:2001fn} to fail badly for large values of the
transverse momentum. Also,  the structure of the dependence of this
process on the various scales which
characterize it (the heavy quark masses, the Higgs mass, and
transverse momentum) is non-trivial and the object of ongoing
investigations~\cite{Banfi:2013eda,Bagnaschi:2015bop}. We expect our results,
though partial, to help in shedding light on these issues, and work on
this is currently ongoing.
\\

{\bf Acknowledgements:} We are grateful to S.~Marzani for innumerable
enlightening discussions and for raising the issue of the relation to
transverse momentum resummation, and to F.~Caola and G.~Zanderighi for
useful comments. We thank R.~D.~Ball, S.~Marzani and especially
F.~Caola for a critical reading of the manuscript.
This work is supported in part by an Italian PRIN2010 grant and
by the Executive Research Agency (REA) of the European
Commission under the Grant Agreement PITN-GA-2012-316704 (HiggsTools).

\appendix
\section{The Higgs $\pt$-impact factor in the $\mtop \to \infty$ limit}
\label{app:Higgsimpact}
We provide here details on the  computation of the double Mellin transform integral
Eq.~\eqref{eq:I} which leads to the final expression of the impact
factor.
We first change variables from $\xi_2$ to a new variable $u$, defined
implicitly as
\beq
\xi_2=1+\xi_1-2\sqrt{\xi_1}u
\eeq
in terms of which, Eq.~\eqref{eq:I} becomes
\begin{align}\label{eq:I2}
I\(M_1,M_2\)=&M_1 M_2 R\(M_1\)R\(M_2\)\notag\\
&\int_0^\infty d\xi_1\,4\xi_1^{M_1}\int_{-1}^{1}du\,\(1-\frac{1}{\sqrt{\xi_1}}u\)^2\(1+\xi_1\)^{M_2-2}\frac{\(1-\sqrt{r}u\)^{M_2-2}}{\sqrt{1-u^2}},
\end{align}
where $r\equiv\frac{4\xi_1}{\(1+\xi_1\)^2}$. 

With straightforward manipulations, Eq.~\eqref{eq:I2} can be rewritten
in terms of a single integral function
\beq
F\(M_1,M_2\)=\int_0^\infty d\xi_1\,\xi_1^{M_1}\(1+\xi_1\)^{M_2}\int_{-1}^{1}du\,\frac{\(1-\sqrt{r}u\)^{M_2}}{\sqrt{1-u^2}},
\eeq
as
\begin{align}
\label{eq:ItoF}
I\(M_1,M_2\)=&M_1M_2 R\(M_1\)R\(M_2\)\notag\\
&\Big[F\(M_1-2,M_2-2\)-2F\(M_1-1,M_2-2\)+F\(M_1,M_2-2\)\notag\\
&-2F\(M_1-2,M_2-1\)+2F\(M_1-1,M_2-1\)+F\(M_1-2,M_2\)\Big].
\end{align}

We compute $F$ by expanding  $\(1-\sqrt{r} u\)^{M_2}$ in powers
of $u$, with the result
\begin{align}
F\(M_1,M_2\)=&\int_0^\infty d\xi_1\,\xi_1^{M_1}\(1+\xi_1\)^{M_2}\int_{-1}^{1}du\,\frac{\(1-\sqrt{r}u\)^{M_2}}{\sqrt{1-u^2}}\notag\\
=&\sum_{k=0}^{\infty}\binom{M_2}{2k}\int_0^{\infty}d\xi_1\,\xi_1^{M_1}\(1+\xi_1\)^{M_2}r^k\int_{-1}^{1}du\,\frac{u^{2k}}{\sqrt{1-u^2}}\notag\\
=&\sum_{k=0}^{\infty}\binom{M_2}{2k}\frac{4^{k}\sqrt{\pi}\Gamma\(\frac{1}{2}+k\)}{\Gamma\(1+k\)}\int_0^\infty d\xi_1\,\xi_1^{M_1+k}\(1+\xi_1\)^{M_2-2k}\notag\\
=&\sum_{k=0}^{\infty}\binom{M_2}{2k}\frac{4^{k}\sqrt{\pi}\Gamma\(\frac{1}{2}+k\)}{\Gamma\(1+k\)}\frac{\Gamma\(1+k+M_1\)\Gamma\(k-1-M_1-M_2\)}{\Gamma\(2k-M_2\)}.
\end{align}
The sum can then be performed in closed form:
\beq
F\(M_1,M_2\)=\frac{\pi\Gamma\(1+M_1\)\Gamma\(1+M_2\)\Gamma\(-1-M_1-M_2\)}{\Gamma\(-M_1\)\Gamma\(-M_2\)\Gamma\(2+M_1+M_2\)}.
\eeq
Substituting this expression in Eq.~\eqref{eq:ItoF} and
 exploiting the properties of the Euler Gamma function  we
 finally get
\beq\label{eq:ires}
I\(M_1,M_2\)=R\(M_1\)R\(M_2\)\left[\frac{\Gamma\(1+M_1\)\Gamma\(1+M_2\)\Gamma\(2-M_1-M_2\)}{\Gamma\(2-M_1\)\Gamma\(2-M_2\)\Gamma\(M_1+M_2\)}\(1+\frac{2M_1M_2}{1-M_1-M_2}\)\right].
\eeq

\end{document}